\documentclass[article, twocolumn, prx, reprint, amsfonts, amsmath, amssymb, superscriptaddress, longbibliography, floatfix]{revtex4-2}
\usepackage[T1]{fontenc}
\usepackage{graphicx}
\usepackage[dvipsnames]{xcolor}
\usepackage[colorlinks,linkcolor=Blue,citecolor=Blue]{hyperref}
\usepackage{braket}
\usepackage{wasysym}
\usepackage{MnSymbol}
\usepackage{bbm}
\usepackage{mathrsfs}
\usepackage{amsthm}
\usepackage{multirow}
\usepackage[dvipsnames]{xcolor}
\usepackage[bottom]{footmisc}
\usepackage[final]{showlabels}

\newcommand{\comment}[1]{}

\newcommand{\eqnref}[1]{Eq.~(\ref{#1})}
\newcommand{\figref}[1]{Fig.~\ref{#1}}
\newcommand{\sfigref}[2]{Fig.~\hyperref[#1]{\ref{#1}#2}}

\newcommand{\secref}[1]{Sec.~\ref{#1}}
\newcommand{\thmref}[1]{Lemma~\ref{#1}}
\newcommand{\appref}[1]{Appendix~\ref{#1}}

\theoremstyle{definition}

\newtheorem{theorem}{Theorem}
\newtheorem{lemma}[theorem]{Lemma}

\newenvironment{sketch}{%
  \proof}{\endproof}

\begin{document}

\title{A numerical study of the {zeros} of the grand partition function of {$k$-mers} on {strips} of width $k$}

\author{Soumyadeep Sarma}
\thanks{\href{mailto:ssoumyadeep@iisc.ac.in}{ssoumyadeep@iisc.ac.in}}
\affiliation{Undergraduate Program, Indian Institute of Science, 560012, Bangalore}


\date{\today}

\begin{abstract}
We study {numerically}, the distribution of the {zeros} of the grand partition function of $k$-mers on a $k \times L$ strip in the complex {fugacity} ({z}) plane. {Using transfer matrix methods}, we find that our results match the {analytical predictions} of Heilmann and Leib for $k = 2$. {However, for $k = 3$, the zeros are confined within a bounded region, suggesting a fundamental difference in critical behavior}. {This indicates that trimers belong to a distinct universality class in some finite geometries. We observe that the density of zeros along multiple line segments in the complex plane reveals a richer structure than in the dimer case.} {Our findings emphasize the role of geometric constraints in shaping the statistical mechanics of $k$-mer models and set the stage for further studies in higher-dimensional lattices.}

\end{abstract}

\maketitle


\section{\label{sec:level1}Motivation and history}

{Phase transitions have been a central topic in statistical mechanics, with key developments such as the mean-field theory, thermodynamic free energy landscapes, classification schemes, and the renormalization group~\cite{vanderwaals1873, maxwell1875, ehrenfest1933, onsager1944, kadanoff1966, wilson1971}. These approaches successfully described critical behavior and universality but often relied on approximations or phenomenological arguments. A more fundamental, model-independent characterization of phase transitions was needed—one that directly captured the emergence of singularities in thermodynamic quantities. This was achieved through the pioneering work of Lee and Yang, who introduced an essential approach based on analyzing the zeros of the partition function in the complex fugacity plane~\cite{leeyang1, leeyang2}}. By studying the grand canonical partition function of the Ising model, they demonstrated that phase transitions could be characterized by the accumulation of zeros along lines in the complex fugacity plane. The Lee-Yang theorem showed that for the ferromagnetic Ising model with real interactions, all partition function zeros lie on the unit circle in the complex fugacity plane, and the density of zeros near the real axis determines the nature of the phase transition. This method provided a powerful tool to probe critical behavior, as the distribution of zeros identifies critical points and characterizes different phases, providing an alternative to mean-field descriptions and demonstrating how non-analytic behavior in thermodynamic quantities arises in statistical models. These results have since been generalized to many models, making partition function zeros central in studies of phase transitions, including lattice gases, spin systems, and percolation models \cite{fisher, griffiths1969, temperley}.

Fisher \cite{fisher} extended the Lee-Yang framework to lattice gas models, which are equivalent to the Ising model in many respects. He demonstrated that partition function zeros for lattice gases exhibit similar properties, allowing for phase transition characterization such as condensation. He also discussed the Lee-Yang edge singularity as a universal critical phenomenon \cite{fisher3}, where the singularity at the edge of the partition function zero distribution acts as a critical point. {W}ork by Baxter and Temperley has focused on models like the hard hexagon and hard square models, where exact solutions provide valuable insights into critical phenomena. {Baxter's exact solution of the hard hexagon model} \cite{baxter} {revealed} the nature of phase transitions in low-dimensional systems with hard-core interactions, {provided further evidence of the universality of partition function zeros in characterizing phase transitions}. Temperley and Lieb \cite{temperley} contributed to understanding two-dimensional lattice systems by analyzing the exact solutions of percolation and dimer models, linking lattice models and algebraic structures, which revealed universality and critical behavior. {The Temperley-Lieb algebra demonstrated deep mathematical connections between partition function zeros, exactly solvable models, and conformal invariance. More recently, numerical approaches~\cite{Assis_Jacobsen_Jensen_Maillard_McCoy_2013} have enabled the analysis of partition function zeros for non-integrable models, extending Lee-Yang's original ideas into new domains}.

{In parallel, the study of $k$-mers (linear molecules composed of k monomers) has gained prominence in statistical mechanics and condensed matter physics due to their relevance in modeling various physical systems~\cite{degennes1993,flory1953,whitesides2002,tobacomosaic,mogilner2003,nanotube,livolant1996,kasteleyn1961,frenkel1985,mccoy2010,nisbet2015}. The adsorption and phase behavior of $k$-mers on lattices serve as fundamental models for understanding liquid crystals \cite{degennes1993}, polymer systems \cite{flory1953}, and molecular self-assembly \cite{whitesides2002}. In biological systems, $k$-mer models describe the self-organization of tobacco mosaic viruses \cite{tobacomosaic} and the behavior of cytoskeletal filaments \cite{mogilner2003}. Carbon nanotube nematic gels \cite{nanotube} and liquid crystalline phases of DNA \cite{livolant1996} have also been successfully modeled using $k$-mers. From a mathematical standpoint, the statistical mechanics of $k$-mers extend the classical dimer model \cite{kasteleyn1961} and exhibit intricate combinatorial structures, which have been explored using Monte Carlo simulations \cite{frenkel1985} and transfer matrix techniques \cite{mccoy2010}. The partition function zeros approach has proven particularly useful for characterizing phase transitions in $k$-mer systems, as demonstrated in recent studies \cite{nisbet2015}. The universality of these results underscores the significance of Lee-Yang's theory beyond spin models, solidifying its role in modern statistical mechanics and interdisciplinary applications.}

In our work, we use the transfer matrix technique to {numerically} generate the polynomial {grand partition} functions {(we drop the word ``grand'' henceforth, and use just ``partition function'' everywhere)} on $k\times L$ strips for $k = 2,3$ for $L$ varying from 1 to 500. We consider the problem of studying the numerical structure of {zeros} of the partition function for $k$-mers on a $k \times L$ lattice ({or a strip of width $k$ and length $L$}) with $k = 2,3$. Recursion relation{s} have been highly powerful for studying such problems and {are} the main tool utilized in this paper. We find that for trimers ($k = 3$) on a $3 \times L$ strip, the partition function zeros are bounded in a region of {fugacity} ($z$) with $-17.34 \leq \mathcal{R}e(z) \leq 0$, as compared to dimers ($k=2$) on 2D lattice structures which are unbounded~\cite{Heilmann_Lieb_1972}. {In our case, we study the configuration of dimers on the $1 \times L$ and $2 \times L$ strips}. We utilize a novel method involving the phase discontinuity of the term $\ln(\lambda_{max}(z))$ across branch cuts to analyze the power law relations of {the} density of {zeros} for such systems. These {lines} of {zeros} are found by {analyzing} the absolute ratio of the two largest eigenvalues of the transfer matrix, and then a general gradient-based algorithm can be utilized to check the distance of the ratio from 1. These bounded zeros also give rise to possible corrections in the density of filled sites in the lattice as a function of real positive $z$, far away from the bounded region of zeros.

The paper is organized as follows: In \secref{sec:sec2}, a{fter explaining some notation}, we discuss the simplest case of a dimer on the $1\times L$ strip ({cf.} \secref{sec:level2}) and extend the model in the scenario of the dimers on a $2\times L$ strip ({cf.} \secref{sec:level2_2}), with the final results and the plots highlighted in \secref{sec:level2_3}. In \secref{sec:sec3}, we look at the $k = 3$ case, highlighting the methods utilized for studying the finite $L$ ({cf.} \secref{sec:level3_1}) and the corresponding Mathematica plots and results ({cf.} \secref{sec:level3_2}). \secref{sec:level3_3} discusses a novel way of finding the {zeros} in the thermodynamic limit ($L \to \infty$) and highlights these in the complex plane. \secref{sec:sec4} summarizes the work, and discusses some open questions regarding this system. Acknowledgments and Appendices follow after that. We discuss the transfer matrix and branch cut calculations in the appendices.

\section{Basic Terminology and transfer matrix method} \label{sec:sec-1}

{We define a $k$-mer as a connected sequence of $k$ sites in a straight line horizontally or vertically. In a configuration with multiple such $k$-mers, a single lattice site may only be occupied by one $k$-mer at a time, or none at all. No two $k$-mers are allowed to overlap on a single site. In the text, a configuration of $k$-mers is a specific placement of $k$-mers on the whole lattice, while an arrangement generally refers to a placement of $k$-mers on some fixed sites. Note that the configuration of an arrangement then refers to a placement of $k$-mers on the whole lattice satisfying the initial arrangement of $k$-mers on some specified sites.}  We define $Z_{L,{k}}({z},h_1,h_2,... h_{k'})$ as the partition function on a strip {with $k'$ rows (or having width $k'$)} {as a function of the {fugacity} $z$, {which is the main parameter}, associated with placing a $k$-mer on the lattice. $z = e^{\beta \mu}$, where $\mu$ is the chemical potential, and $\beta = 1/T$ is the inverse temperature. Chemical potential and temperature for lattice models are more generally absorbed in the {fugacity} parameter $z$ as they are not defined in the conventional sense here~\cite{huang1987,landau1980,pathria2011,fowler1939,yeomans1992,hill1962}. The} corresponding boundary conditions {are} defined by $\{h_1,h_2,...h_{k'}\}$, where $0\leq h_i \leq k - 1$ are the number of lattice sites occupied by $k$-mer {placed} on row $i$ and column $L$, to the right {of the column}. {Note that the notation $Z_{L,k}(z,h_1,h_2,\cdots,h_{k'})$ does not by itself bound the $k$-mers to be placed inside a $k' \times L$ strip, allowing an extended boundary beyond even the rightmost column $L$ for our strip}. {For our results, under this notation, if we take a strip of length $L$, we are interested in the {zeros} of $Z_{L,{k}}({z},0,0,\cdots)$ as there are only $L$ columns, and a $k$-mer extending beyond the rightmost column is not a valid arrangement considered by the partition function in a strip with $L$ columns.} As a simple example, if we denote {$Z_{L}(z):= Z_{L,{k}}({z},0)$} as the partition function for a $k$-mer on a $1\times L$ strip, with the {fugacity} of each $k$-mer being $z$, then it is straightforward to derive the recursion relation:
\begin{eqnarray}
\label{eqn:eqn1}
    {Z_{L}(z) = Z_{L-1}(z) + zZ_{L-k}(z).}
\end{eqnarray}
{More generally, we use the notation $Z_{L,{k}}(z):=Z_L({z},0,0,\cdots)$ for the partition function of $k$-mers on $k'=k$ rows (except for $k$-mers on the $1 \times L$ strip, where we don't put a subscript $k$)}. {Later in the text, when we write the partition function as a linear combination of eigenvalues to the power of the strip length, we use $\tilde Z_{L,k}(z)$ as notation for truncating the sum to only the largest few eigenvalues.} {In the following subsections, we highlight how the transfer matrix method can be used to rewrite such recursion relations into a matrix equation in the case of dimers and thus used to obtain the partition function analytically or numerically.}

\label{sec:sec2}

\subsection{Dimer on $1\times L$ strip}
\label{sec:level2}
We set k = 2 in \eqnref{eqn:eqn1}. Writing it in matrix form gives us (we omit the ${z}$ from the brackets, as it is implied):
\begin{eqnarray}
    \begin{bmatrix}
           {Z}_{L+1} \\
           {Z}_{L} 
         \end{bmatrix} &= \begin{pmatrix}
           1 & z \\
           1 & 0 \\
         \end{pmatrix}
         \begin{bmatrix}
             {Z}_{L} \\
             {Z}_{L-1} 
         \end{bmatrix}.
\end{eqnarray}
Clearly, if we label the square matrix as $T$, we see that it is $T$ multiplied repeatedly on an initial column vector, which is of the form:
\begin{eqnarray}
   \begin{pmatrix}
           {Z}_{L+1} \\
           {Z}_{L} 
         \end{pmatrix} &= T^{L-1}
         \begin{pmatrix}
             {Z}_{2} \\
             {Z}_{1} 
         \end{pmatrix}.
\end{eqnarray}
We can easily find the partition functions ${Z}_{2}$ and ${Z}_{1}$. For $L = 1$, there is only one site, and no dimer can be placed, thus ${Z}_{1} = 1$, whereas for $L=2$, either there is no dimer or there is one dimer, implying that ${Z}_{2} = 1+z$. Using spectral decomposition, we can write $T^{L-1} = UD^{L-1}U^T$, where D is a diagonal matrix of eigenvalues ({$\lambda_{\pm}$}) of $T$, {exponentiating} which gives:
\begin{eqnarray}
  T^{L-1} &= U
         \begin{bmatrix}
             \lambda_{{+}} & 0 \\
             0 & \lambda_{{-}} 
         \end{bmatrix}^{L-1}U^T,\\
         &=  U
         \begin{bmatrix}
             \lambda_{{+}}^{L-1} & 0 \\
             0 & \lambda_{{-}}^{L-1} 
         \end{bmatrix}U^T.
\end{eqnarray}
We see that the matrix equation seems to imply that ${Z_{L}}$ will have a form that is a linear combination of eigenvalues raised to the power $L$:
\begin{eqnarray}
  {Z_{L}(z)} = c_1\lambda_{{+}}(z)^L + c_2\lambda_{{-}}(z)^L.
\end{eqnarray}
The eigenvalues for the $T$ matrix are $\lambda_{{\pm}} = \frac{1 \pm \sqrt{1+4z} }{2}$. To determine the constants, we use the two values of ${Z_{L}(z)}$ at $L = 1$ and $L = 2$ and solve for the obtained linear equations to give:
\begin{eqnarray}
  {Z_{L}(z)} = \frac{1}{2}\left(\frac{1 + \sqrt{1+4z} }{2}\right)^L + \frac{1}{2}\left(\frac{1 - \sqrt{1+4z} }{2}\right)^L.
\end{eqnarray}
The {zeros} of this partition function can be arrived at {by} solving the equation:
\begin{eqnarray}
  \left(\frac{1+\sqrt{1+4z}}{1-\sqrt{1+4z}}\right)^L = -1,
\end{eqnarray}
which becomes,
\begin{eqnarray}
\label{eqn:eq9}
  \frac{1+\sqrt{1+4z_r}}{1-\sqrt{1+4z_r}} = e^{\frac{i\pi(2r+1)}{L}},
\end{eqnarray}
with $r$ going from $0$ to $L-1$, $z_r$ is the $r$-th root of ${Z_L}$. The L.H.S has mod 1 only if $1+4z < 0$, implying that {zeros} lie on the negative real axis. {We can define the fractional density of zeros $\phi_L(z)$ for a strip of length $L$ as:
\begin{align}
    \int_{z_0}^{z_r}\phi_L({t}){dt} = \frac{r}{L}
\end{align}
We take $\phi(z) := \lim_{L \to \infty} \phi_L(z)$. Now,} note that from \eqnref{eqn:eq9}:
\begin{align}
z_r &= -\frac{1}{4} - \tan^2(\frac{\pi}{2L}(2r+1)),\\
\label{eqn:r}
r &= \frac{L}{\pi}\tan^{-1}(\sqrt{-(z_r + 1/4)}). 
\end{align}
Thus, we have from \eqnref{eqn:r}:
\begin{align}
   \int_{z_0}^z \phi (t) dt &= \frac{1}{\pi} \tan^{-1}(\sqrt{z_0 - z}),\\
   \phi(z) &= \frac{1}{2\pi \sqrt{z_0 - z}} \left(\frac{1}{1 + (z_0 - z)} \right).
\end{align}
This density function, for values close to $z_0$, is $\approx |z-z_0|^c$, with $c = -0.5$, which is the critical exponent, as defined in Lee-Yang's work \cite{leeyang1} (or $c = 0.5$, if we take the power law to go like $1/|z-z_0|^c$). It diverges at the endpoint of the line of {zeros}, i.e., at $z_0 = \frac{-1}{4}$, and goes to zero as $z \to \infty$ (at large negative $z$, the function goes as $1/|z_0 - z|^{3/2}$).

\subsection{Dimer on a ladder i.e. $2\times L$ strip}
\label{sec:level2_2}

\begin{figure*}

\includegraphics[width = 1.5\columnwidth]{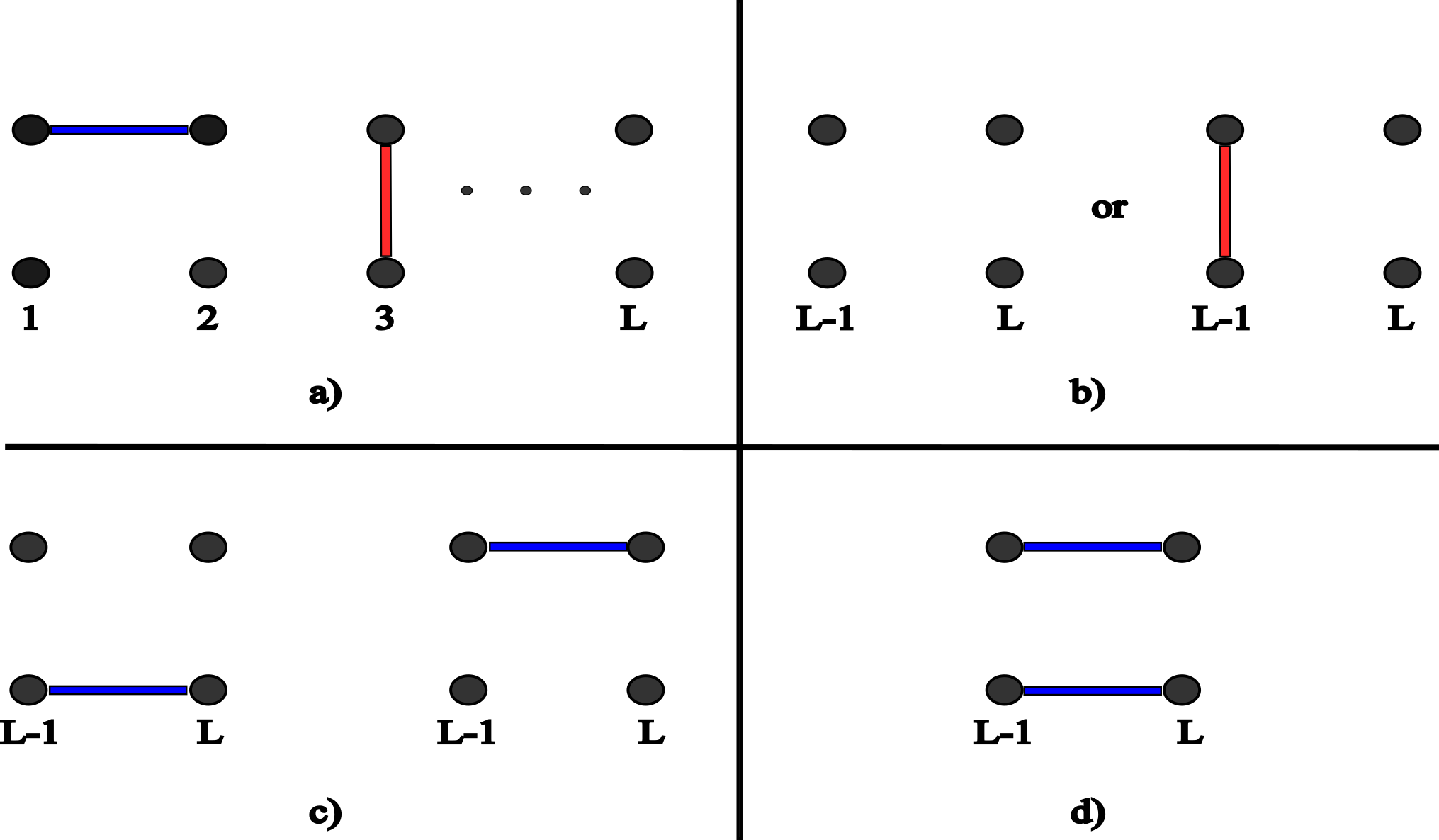}
\caption{{The arrangement of dimers on the last two columns in a $2 \times L$ strip.} a) The numbering of columns in the strip. b) No dimers going across the last two columns {and no dimer on last column}. c) One dimer going across the last two columns. c) Two dimers going across the last two columns }
\label{fig:fig_1}
\end{figure*}

{We can number the columns with the label $x$ going from 1 to $L$. As seen in \figref{fig:fig_1}}, the possible {arrangements} for horizontal dimers across the last two columns for a strip of length $L$ are:

\begin{itemize}
    \item No dimer goes across column $x = L-1$ and $x = L$, {and no vertical dimer on column $x = L$.}
    \item Exactly one dimer goes across $x = L-1$ and $x = L$.
    \item Two dimers go across $x = L-1$ and $x = L$.
\end{itemize}

We label the sum {of configurations} over each such {arrangement} (or partition function for each such {arrangement}) as $Z_1(L),~Z_2(L),$ {and} $Z_3(L)$. {This notation of labeling partition functions based on number of $k$-mers going across the last $k$ columns will be utilized again for the trimers on a $3 \times L$ strip case in Sec.~\ref{sec:level3_1}.} The following recursion relations are obtained:
\begin{align}
  Z_1(L) &= (1+z)Z_1(L-1) + Z_2(L-1) + Z_3(L-1),\\
    Z_2(L) &= 2zZ_1(L-1) + zZ_2(L-1),\\
    Z_3(L) &= z^2Z_1(L-1). 
\end{align}

We can motivate these recursion relations as follows: For the first relation, clearly, an additional vertical dimer can be added in the $x = L-1$ column in addition to not adding anything. For the second relation, two horizontal dimers can go across in the case of $Z_1(L)$ and one horizontal dimer in the case of $Z_2(L-1)$. For the third relation, the only scenario possible is two horizontal dimers going across for the case $Z_1(L-1)$. The corresponding matrix equation is:
\begin{eqnarray}
    \begin{bmatrix}
           Z_1(L) \\
           Z_2(L)\\
           Z_3(L)
         \end{bmatrix} &= \begin{pmatrix}
           1+z & 1 & 1 \\
           2z & z & 0 \\
           z^2 & 0 & 0
         \end{pmatrix}
         \begin{bmatrix}
              Z_1(L-1) \\
              Z_2(L-1)\\
              Z_3(L-1) 
         \end{bmatrix},
\end{eqnarray}
which, using the simple recursive idea as present in \secref{sec:level2}, becomes:
\begin{eqnarray}
\label{eqn:eq14}
    \begin{bmatrix}
           Z_1(L) \\
           Z_2(L)\\
           Z_3(L)
         \end{bmatrix} &= \begin{pmatrix}
           1+z & 1 & 1 \\
           2z & z & 0 \\
           z^2 & 0 & 0
         \end{pmatrix}^{L-1}
         \begin{bmatrix}
              Z_1(2) \\
              Z_2(2)\\
              Z_3(2) 
         \end{bmatrix}.
\end{eqnarray}
Clearly, ${Z_{L,{2}}({z},0,0):=}~{Z}_{L,{2}}(z) = Z_1(L+1)$ and $Z_1(2) = 1+z$, $Z_2(2) = 2z$, $Z_3(2) = z^2$ . Using the same spectral decomposition ideas as seen previously, we arrive at: 
\begin{eqnarray} \label{eq:someq}
  {Z_{L,{2}}(z)} = c_1\lambda_1(z)^L + c_2\lambda_2(z)^L+ c_3\lambda_3(z)^L.
\end{eqnarray}
While the cubic equation can be solved exactly in closed form, the expressions for the roots are rather complicated, and omitted here, which also does not allow us to do the same analysis as in \eqnref{eqn:eq9}. It is quite straightforward to calculate powers of $T$ directly by direct multiplication. We wrote a Mathematica code that evaluates the polynomials $Z_i(L)$ explicitly, and {determines} the {zeros} numerically for $L$ up to  200. The relation was explicitly checked by drawing configurations for $L = 2,3,4$ and the notes and the corresponding codes on this have been uploaded on GitHub (see Acknowledgements for the link)\\

\subsection{{Numerical results}: {Zeros} of the partition function}

\label{sec:level2_3}

As mentioned in the last section, for the $2\times L$ strip, it is much easier to work with the {t}ransfer matrix than the analytic expressions of the eigenvalues. The code for an input $L$, outputs the form of the partition function and its {zeros} and plots them in the complex plane.

\begin{figure*}
     \centering
        \includegraphics[width=2\columnwidth]{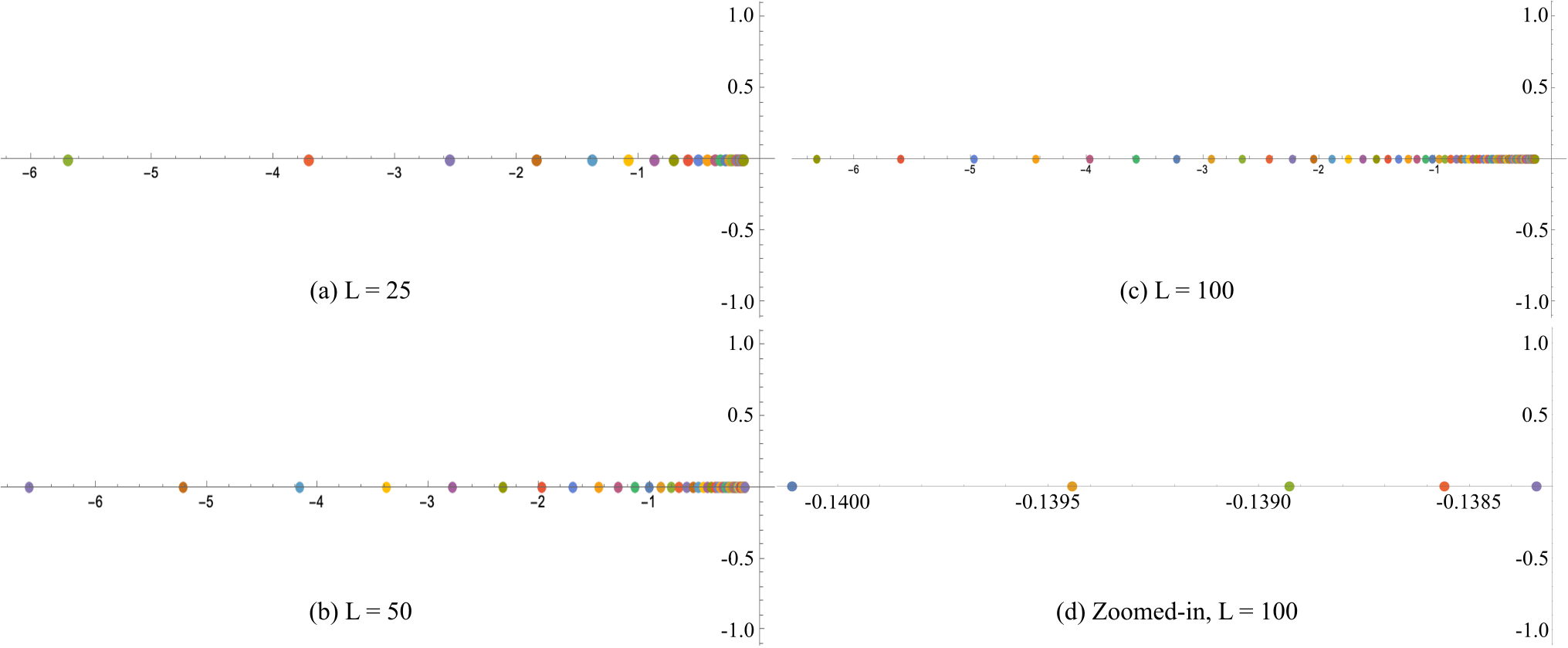}
        \caption{Plots of {zeros} of the partition function {in the case of dimers on a $2 \times L$ strip (i.e. dimers on a ladder-like lattice) for (a) $L = 25$, (b) $L = 50$, (c) $L = 100$. With increasing $L$, we see the density of zeros increasing near the rightmost critical point (the point closest to the origin, from where all the zeros seem to "start" and go till $-\infty$) $z_0 = -0.13826671$ as shown in (d) for $L = 100$.}}
        \label{fig:fig_2}
\end{figure*}

The results are what was expected: for any $L$, the {zeros} lie on the negative real axis, while the density of {zeros} approaches higher and higher values at one point, i.e., all the {zeros} seem to congregate at that point. The plots are given in \figref{fig:fig_2}(a,b,c) for $L = 25, {5}0, \text{ and } 100$. This is in perfect agreement with the theorem by Heilmann and Lieb applicable for any graph~\cite{Heilmann_Lieb_1972}, which says that the {zeros} of the partitions function of dimers on graphs always lie on the negative real axis. Our data is consistent with this result. 

{Here, we present an idea as to how we can analytically find the endpoints of a line of zeros. From Eq.~\eqref{eq:someq}, if we truncate the sum to the first two largest eigenvalues (taken $|\lambda_1(z)| > |\lambda_2(z)| > |\lambda_3(z)|$ with the coefficients $c_i$ absorbed in the eigenvalues, tilde represents the truncated expression, $\tilde Z_{L,{2}}(z):= \tilde Z_{L,{2}}({z},0,0)$, as introduced in Sec.~\ref{sec:sec-1}):
\begin{align}
    \tilde{{Z}}_{L,{2}}(z) = \lambda_1 (z)^L + \lambda_2 (z)^L.
\end{align}
For large $L$, this function is a very close approximation to the actual partition function. Hence, we can just equivalently find roots of $\tilde{{Z}}_{L,{2}}(z)$, which will just give:
\begin{align}
    &\left(\frac{\lambda_1 (z)}{\lambda_2 (z)}\right)^L = -1, \\ \label{eq:theta}
    \frac{\lambda_1 (z)}{\lambda_2 (z)} = &e^{i\theta}, \text{  
 where } \theta = \frac{(2m+1)\pi}{L}.
\end{align}
Now, note that the endpoints of the line of {zeros} are where $\theta \to 0$, i.e. $\lambda_1(z_0) \to \lambda_2(z_0)$, which can be found by considering the common roots to the characteristic equation (obtained from transfer matrix) and its first differential at a specific $z = z_0$:
\begin{align} \label{eq:cop}
    f(\lambda,z_0) = 0 \quad \text{and} \quad \frac{ \partial f}{\partial \lambda}\bigg \vert_{z = z_0} = 0.
\end{align}
Then these are coupled polynomial equations in $\lambda$ and $z_0$. In coupled polynomial equations, we can eliminate one variable, and get a polynomial equation in the second variable only, which is $z_0$ in this case. Solving this leads to calculating the endpoints up to a high precision.
}

{Let us try to analytically ascertain the rightmost critical point $z_0$ for the case of dimers on $2 \times L$ strip. We use the method of finding common roots of the characteristic equation of the transfer matrix by solving coupled polynomial equations. The reasoning behind this idea is explained in more depth in Sec.~\ref{sec:level3_3} (cf. Eq.~\eqref{eq:cop}). From the transfer matrix in Eq.~\eqref{eqn:eq14}, we have the characteristic equation and its first derivative:
\begin{align}
\label{eqn:rootz} 
    f(\lambda,z) &= \lambda^3 + \lambda^2(1 + 2z) - \lambda z + z^3  \\ \label{eq:part}
    \frac{\partial f}{\partial \lambda} &= 3\lambda^2 + 2\lambda(1 + 2z) - z
\end{align}
At the critical point $z = z_0$, we have that $f(\lambda,z_0) = (\partial f/\partial \lambda)_{z_0} = 0$. Multiplying Eq.~\eqref{eq:part} with $\lambda/3$ and subtracting from Eq.~\eqref{eqn:rootz} gives us another quadratic polynomial in $\lambda$. Substituting for $\lambda^2$ then gives us a linear equation in $\lambda$ given by:}
\begin{align} \label{eqn:linear}
  \lambda = \frac{-9z_0^3 + 4z_0^2 + z_0}{16z_0^2 + 6z_0 + 2}  
\end{align}
{Substituting \eqnref{eqn:linear} in \eqnref{eq:part} and solving numerically for $z_0$ gives us} $z_0 = -0.13826671$, which is in excellent agreement with the numerical results in \figref{fig:fig_2}(d). It is easy to see that near $z_0$, the eigenvalues $\lambda_1$ and $\lambda_2$ obtained from \eqnref{eq:part} have a square root singularity.

\section{Trimers on $3 \times L$ strip}
\label{sec:sec3}
Let us now look at the  $k = 3$ version of the problem. One immediate challenge that we face is the increased number of configurations to consider. However, by utilizing symmetry arguments, we can make the problem much easier to deal with. There are primarily two notations that can be utilized in the transfer matrix method. The first is the standard notation commonly used in statistical mechanics to study lattice problems, while the second is something we found convenient while thinking about the problem, {and as was also introduced in Sec.~\ref{sec:level2_2}.}

\subsection{Transfer Matrix method revisited}
\label{sec:level3_1}

\begin{figure*}
\centering
\hspace*{15mm}
\includegraphics[width = 2\columnwidth]{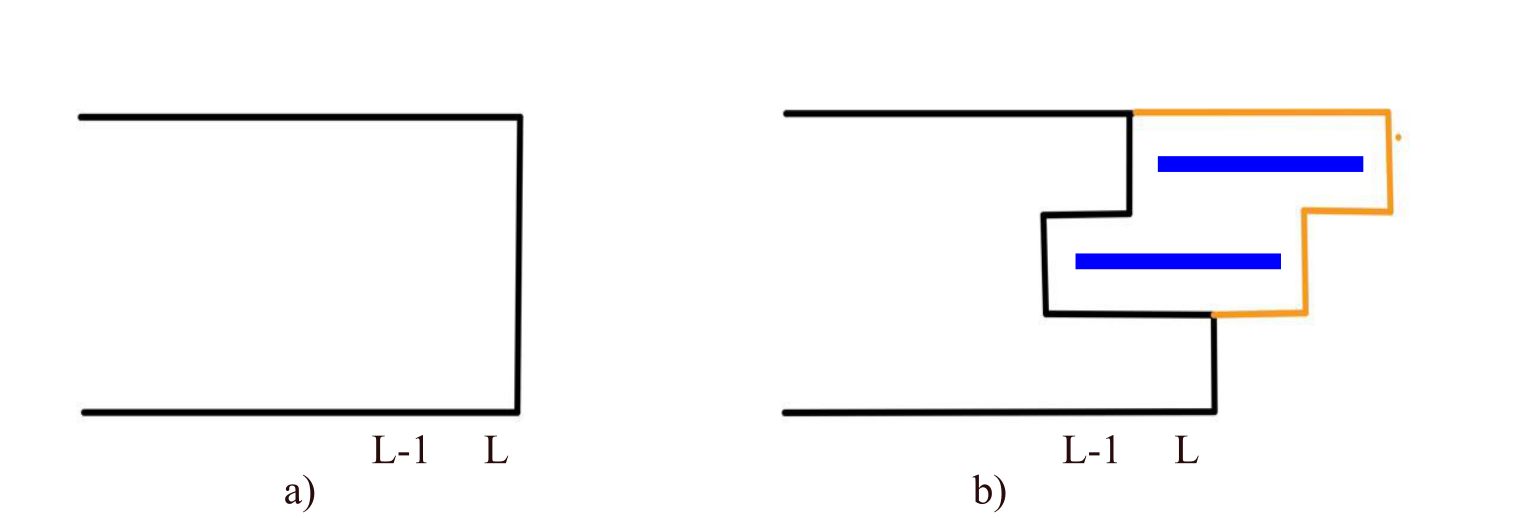}
\caption{Examples of the first notation for partition functions of trimers on $3 \times L$ strip. The orange lines represent the extended boundary, while the black lines represent the actual boundary {inside which trimers in different configurations can be placed.} a) $Z_L(0,0,0)$ {is the partition function containing all configurations of trimers in a $3 \times L$ strip, where the black and orange boundaries overlap.} b) $Z_L(2,1,0)$ {is a partition function involving the fixed extended boundary (orange) going beyond the rightmost column of a strip with length $L$ or $L+1$, while the actual boundary (black) highlights the space present for other trimers to be placed.}}
\label{fig:config_trimer2}
\end{figure*}

\begin{figure*}
\centering
\includegraphics[width = 2\columnwidth]{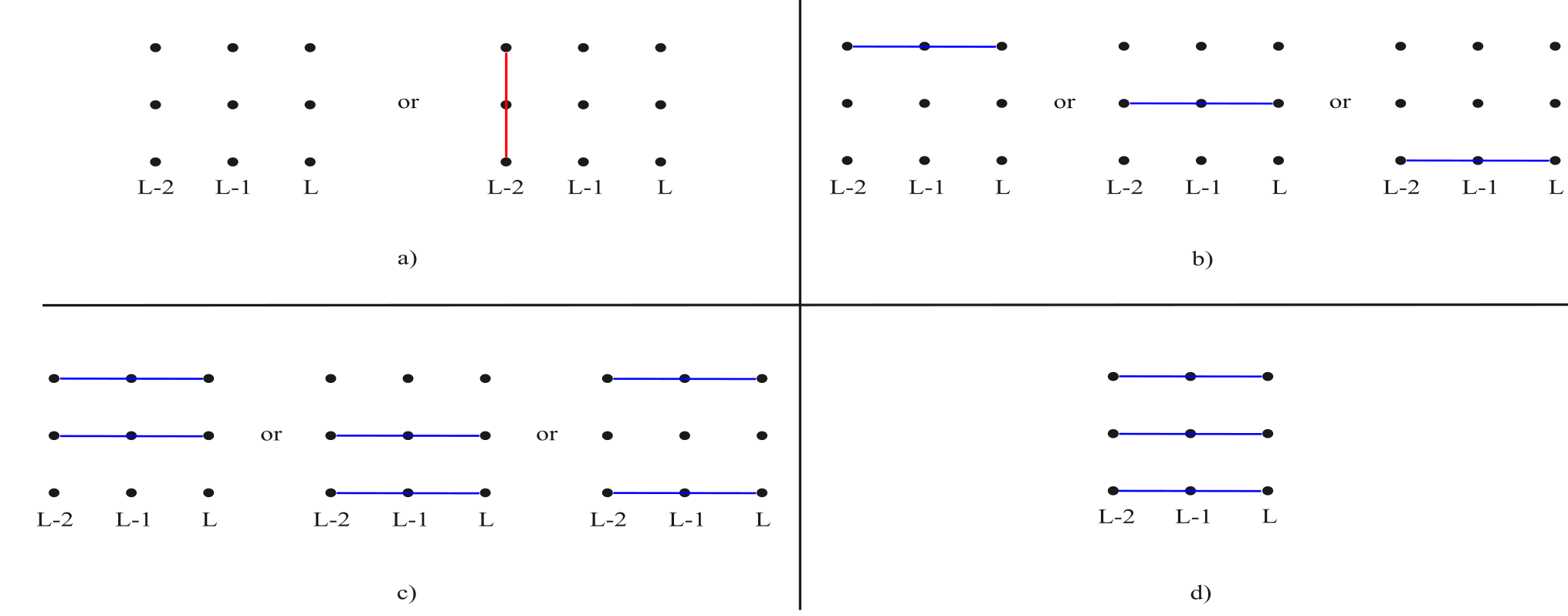}
\caption{Pictorial representation of the second notation for partition functions of trimers on $3 \times L$ strip. a) No trimer going across columns $L-2$ to $L$ {and no vertical trimers on columns $L-1,L$}. b) One trimer going across columns $L-2$ to $L$. c) Two trimers going across columns $L-2$ to $L$. d) Three trimers going across columns $L-2$ to $L$.}
\label{fig:config_trimer1}
\end{figure*}

The setup is as follows: The strips go from left to right (the column numbering from 1 to L is from left to right), with each column having 3 sites. Then we have the following two notations which can be used to establish the transfer matrix recursion relations:

\begin{enumerate}

     \item The {partition function} is denoted as $Z_{L,{3}}({z},h_1,h_2,h_3)$, where $h_i \in \{0,1,2\}$. The $h_i$ denotes how many sites to the right of the site at column $L$ and row $i$ are occupied by horizontal trimers (if 0, it means no sites to the right are occupied. If 1, it means that the horizontal trimer occupies that site, one site to the right and one to the left of row $i$ and column $L$, and 2 means that the site is a tail-end of a horizontal trimer extending to the right from the site on column $L$ and row $i$). We can see that the values $h_1,h_2,h_3$ {fix} the {extended} "rightmost boundary" of {all the possible configurations}. {For a strip of fixed length $L$, clearly we cannot put trimers beyond the rightmost column, and hence we wish to study $Z_{L,{3}}({z},0,0,0)$, as any non-zero value as an argument in $Z_L$ will include configurations where the trimers extend beyond the rightmost column $L$ (cf. \figref{fig:config_trimer2}).} Vertical trimers are directly multiplied with a factor of $z$ in these configurations sums as and when required
     
    \item The {partition functions} are denoted as $Z_i(L)$, $i \in \{1,2,3,4\}$, where: {(a)} $i=1$ means no horizontal trimer crosses from column number $x = L-2$ to $x=L$, {and no vertical trimers on columns $x = L-1,L$}. {(b)} $i=2$ means only one horizontal trimer crosses over from $x = L-2$ to $x=L$. {(c)} $i = 3$ means two horizontal trimers cross over from $x = L-2$ to $x=L$, and {(d)} $i=4$ means three trimers cross over {(cf. \figref{fig:config_trimer1})}.

\end{enumerate}

{\appref{apn:app_1} shows how to transform between the two notations}. The first notation can be generalized to the general $k$-mers of $k \times L$ strip as configurations of the type $Z_{L,{k}}({z},h_1,h_2,h_3....h_k)$, with $h_i \in \{0,1,2,...k-1\}$. A general lemma can be stated here about such configurations:

\begin{lemma}
\label{thm:main_thm}
Let $S$ be the boundary of the strip configuration (arrangement of $\{h_i\}$, $1 \leq i \leq k$), such that at least one value is non-zero, i.e. $\exists i, h_i \neq 0$. Let $Z_{L,{k}}({z},S)$ be the configuration{al} weight associated with $S$. Let $S'$ be a permutation of $S$, such that $S' = \{h_i'\}$. Then $Z_{L,{k}}({z},S)$ = $Z_{L,{k}}({z},S')$.

\end{lemma}

\begin{sketch}
Since at least one $h_i \neq 0$, that implies that there cannot be any configuration with a vertical trimer on column $L$. This means that all configurations {contributing to} $Z_{L,{k}}({z},S)$ have sites on column $L$ that can only be occupied by horizontal trimers, which satisfies the boundary conditions. But since, w.r.t the horizontal trimers, the boundary is symmetric, any permutation of $S$ (i.e. $S'$) also has the same total configuration{al} weight $Z_{L,{k}}({z},S)$, implying that $Z_{L,{k}}({z},S)$ = $Z_{L,{k}}({z},S')$.   
\end{sketch}

\thmref{thm:main_thm} implies that only the {configurations with \textit{unique} boundaries} ($S$) are necessary to find the recursion relation{s}. The problem is similar to filling $k$ options that add up to k (the $k$ options are values between 0 to k-1, which can be chosen with repetition). Using standard combinatorics, this evaluates to a possible $2k-1 \choose k$ configurations for $S$, which is also the length of the column vector in the transfer matrix method. For both methods to be valid, the dimensions of the {t}ransfer {m}atrices should be the same in both cases, as they encode the same problem in different ways. For $k = 3$, there are $3^3$ basic partition functions, but the size of the matrix can be reduced by using these symmetries of the problem. This evaluates to a $10 \times 1$ vector on which a $10 \times 10$ (reduced) matrix acts. The details on the derivation {of the transfer matrix in Eq.~\eqref{eq:tm3}} are listed in \appref{apn:app_1}. We use the second notation to derive the transfer matrix recursion $X_{L+1} = \Gamma (z) X_{L}$, where we have:
%
\begin{equation}
    \begin{aligned} \label{eq:tm3}
    \Gamma (z) &= \begin{pmatrix}
          1 + z & 0 & 0 & 0 &  1 & 1 & 1 & z^2 + 2z & z & 2z^2 \\
          3z & 0 & 0 & 0 &  2z & z & 0 & 4z^2 & 2z^2 & 0 \\
          3z^2 & 0 & 0 & 0 &  z^2 & 0 & 0 & 2z^3 & z^3 & 2z^4 \\
          z^3 & 0 & 0 & 0 &  0 & 0 & 0 & 0 & 0 & 0 \\
          0 & 1 & 0 & 0 &  0 & 0 & 0 & 0 & 0 & 0 \\
          0 & 0 & 1 & 0 &  0 & 0 & 0 & 0 & 0 & 0 \\
          0 & 0 & 0 & 1 &  0 & 0 & 0 & 0 & 0 & 0 \\
          0 & 0 & 0 & 0 &  1 & 0 & 0 & 0 & 0 & 0 \\
          0 & 0 & 0 & 0 &  0 & 1 & 0 & 0 & 0 & 0 \\
          0 & 0 & 0 & 0 &  0 & 0 & 0 & 1 & 0 & 0 \\
         \end{pmatrix}, \\[10pt]
    \end{aligned}
\end{equation}
%
where,
\begin{align} \nonumber
\label{eqn:tm_main}
    X_L = [&Z_1(L+1),  Z_2(L+1),  Z_3(L+1),  Z_4(L+1),  Z_2(L), \notag \\ 
         &Z_3(L),  Z_4(L),  Z_2(L-1),  Z_3(L-1),  Z_2(L-2)]^{T}.  
\end{align}
\subsection{Results and plots of the numerical study}
\label{sec:level3_2}

\begin{figure}
    \centering
    \includegraphics[width=\columnwidth]{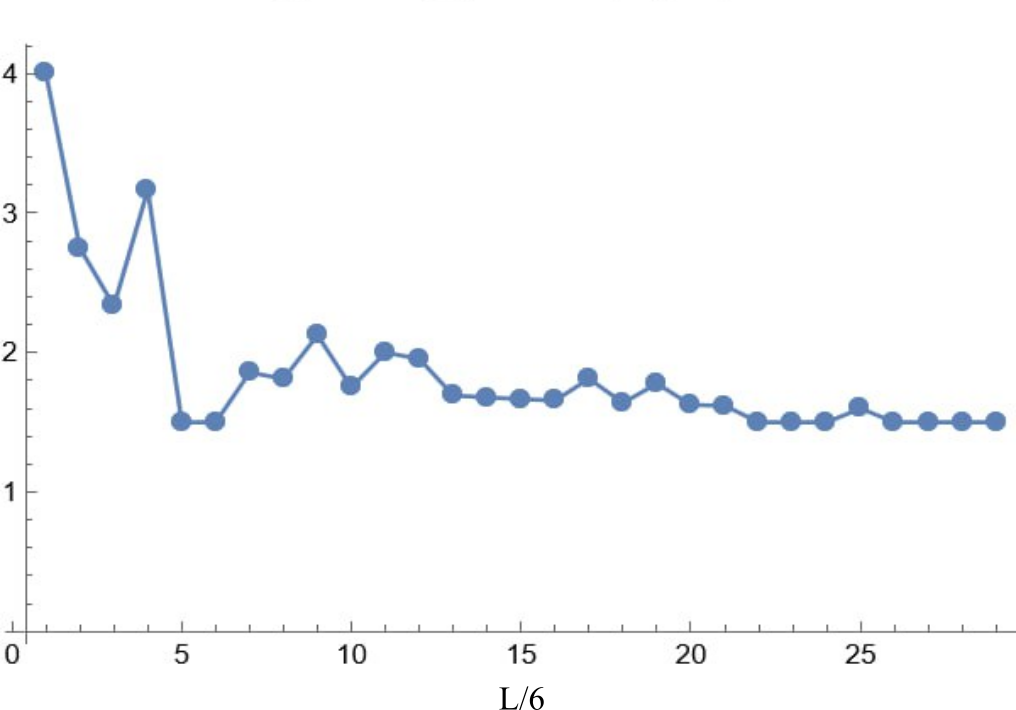}
    \caption{Ratio of real roots to complex roots {versus} $L/6$, {for the partition function of trimers on $3 \times L$ strip,} starting from $L=10$. The value asymptotically reaches $1.5$ for large $L$}
    \label{fig:real_complex}
\end{figure}

\begin{figure}
    \centering
    \includegraphics[width=1\columnwidth]{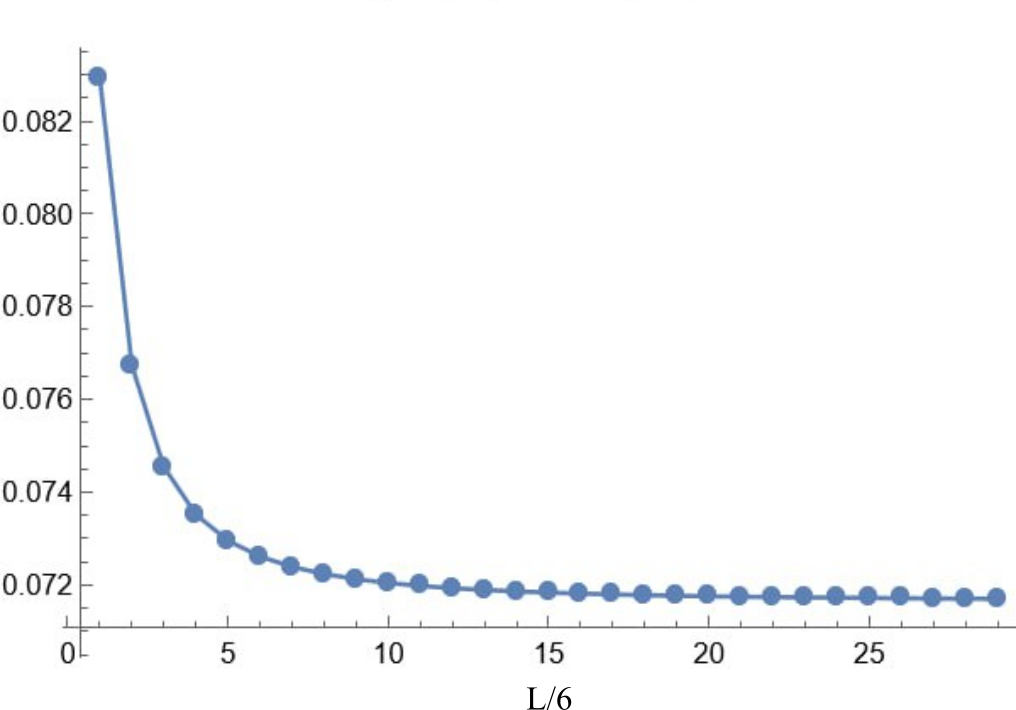}
    \caption{Modulus of smallest negative root {versus} $L/6$, {for the partition function of trimers on $3 \times L$ strip,} starting from $L = 10$ {and reaching asymptotic value $x \approx 0.071$ as $L \to \infty$}}
    \label{fig:small_neg}
\end{figure}

\begin{figure}
    \centering
    \includegraphics[width=\columnwidth]{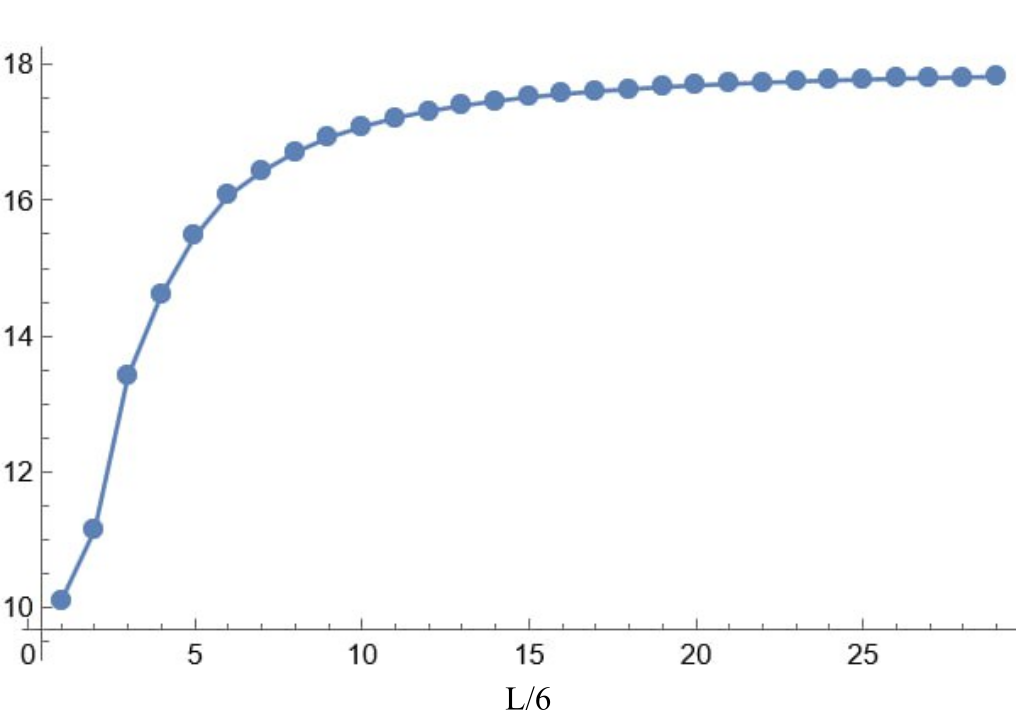}
    \caption{Modulus of largest negative root {versus} $L/6$, {for the partition function of trimers on $3 \times L$ strip,} starting from $L = 10$ {and reaching asymptotic value $c \approx 17.665$ as $L \to \infty$}}
    \label{fig:lar_neg}
\end{figure}

In this section, {we discuss the numerical results obtained for trimers on $3 \times L$ strip.} We consider all plots starting from $L \geq 10$ as otherwise, the system size is too small to be of any significance. Since all roots $z$ are such that $\mathcal{R}e(z) < 0$, we say that the smallest negative root is the one where $\mathcal{R}e(z)$ is closest to zero, and the largest negative root is the one where $Re(z)$ is the farthest from it.

\figref{fig:real_complex} plot{s} the ratio of real roots to imaginary ones as a function of L. We see that the ratio seems to approach an asymptotic value of 1.5 in the thermodynamic limit. \figref{fig:small_neg} and \figref{fig:lar_neg} plot the modul{i} of smallest and largest negative roots as a function of $L$. {In both these figures, we see that the moduli values of the roots reach asymptotic values $x$ and $c$, respectively, in the thermodynamic limit. We observe that both the largest and smallest negative roots lie in the annulus $x \leq |z| \leq c$ , implying the fact that in the thermodynamic limit ($L \rightarrow \infty$), all roots of the partition function are bounded between two values $x$ and $y$, which are $x \approx 0.071$ and $c = 17.665$ (the actual largest negative root is imaginary and always comes in conjugate pairs, with values around $-17.34 \pm i3.36$, such that $c = |-17.34 \pm i3.36| \approx 17.665$), such that for any $z$ to be a root of the partition function $\lim_ {L\to\infty} {Z_L(z)}$, we have $-17.34 \leq Re(z) \leq -0.071$.} This result can be also seen analytically if we consider the characteristic equation of the {transfer} matrix in Eq.~\eqref{eq:tm3} as $f(\lambda) = |T - \lambda I| = 0$ (where $|A|$ means determinant of matrix A). Then we have:
\begin{align*}
    f(\lambda) &= \lambda^{10} - \lambda^9(1+z) -\lambda^8(2z) - \lambda^7(z+5z^2 + 2z^3) \\ \nonumber
    &-\lambda^6 (2z^2 + 2z^3 - z^4) -\lambda^5(5z^3 - z^4) + \lambda^4(2z^4 + z^5 +z^6) \\\nonumber
    &-\lambda^3(10z^5 + 4z^6) - \lambda^2 (8z^6 - 6z^7) + \lambda(2z^8) + 4z^9. \\
\end{align*}
The expression $f(\lambda) = 0$ gives us the eigenvalues of the transfer matrix as a function of $z$. If we put $\lambda = zy$ and divide {t}he entire expression by $z^{10}$ we have:
\begin{align*}
    f(y) = 0 = y^{10} - y^9\left(1+\frac{1}{z}\right) - y^8\left(\frac{2}{z}\right) - y^7\left(2 + \frac{5}{z} + \frac{1}{z^2}\right) + ...
\end{align*}
This equation can be solved for $y$ as a function of $\frac{1}{z}$, and is well-behaved for large $z$. Thus we have that:
\begin{align}
    y(z) &= y_0\left(1 + \frac{A_1}{z} + \frac{A_2}{z^{{2}}} + ...\right)\\
    \lambda (z) &= y_0z\left(1 + \frac{A_1}{z} + \frac{A_2}{z^{{2}}} + ...\right)
\end{align}
{Thus, this} implies that since $z \to \infty$ is not a singular point for the function $\ln(\lambda/z)$, the {zeros} are bounded in a finite region.\\

Clearly, if most of the density of {zeros} is within a distance $R$ of the origin, then analogous to charges on a line, they produce an "electric field" $Q/z$ at $z >> R$, where Q is the "total charge density". For reference, the density of covered sites per total number of sites is obtained from the partition function as:
\begin{align}
    \rho (z) = z \frac{d \ln \lambda}{dz},
\end{align}
where $\lambda$ is the largest eigenvalue of the transfer matrix. It is easily seen that if $\lambda = kz(1 + A_1/z + A_2/z^2 + ...)$, then at $z \to \infty$, the density will be a constant (equal to $Q$, if we go by the general electric charge analogy where $\rho = z \partial V/ \partial z$ with $V \equiv |\ln \lambda|$ as the potential). At large real positive but finite $z$, we can say that the density function takes the Laurent expansion form~\cite{Arfken2011} of $\rho = A + B/z + C/z^2 + \cdots$, which could be understood as:
\begin{align}
    \rho (z) = \rho_\infty - \frac{A}{z^a} + \text{  higher order corrections  }.
\end{align}
This form determines the behavior of density at large real positive but finite $z$. We can also ask how much the density as a function of $z$ deviates from the thermodynamic limit in the case of L being finite. Also, note that since the real part of $\ln(\lambda(z))$ is the electrostatic potential due to the charge distribution. Then, one can calculate the electric field, and the discontinuity in the normal component of the electric field gives the charge density on the line of {zeros}. This is a way to directly determine {charge density} from $\lambda(z)$.

\subsection{Ways of reducing dimension, and other possible approaches}

\label{sec:level3_3}

Let us motivate how to reduce the dimensionality of the transfer matrix method (i.e. reduce the number of independent configurations required as follows). Instead of working with partition functions, we now work with generating functions. Take the generating function $f(x,h_1,h_2,h_3)$ defined as:
\begin{align}
    f(x,h_1,h_2,h_3) = \sum_{L = 0}^\infty x^L {Z}_{L,{3}}({z},h_1,h_2,h_3).
\end{align}
Take $h_1=2,~h_2 = 2 \text{ and } h_3=2$ as an example. Clearly, we see that $Z_{L,{3}} ({z},2,2,2)$ can only give rise to ${Z}_{L+1,{3}} ({z},1,1,1)$, which can only give rise to ${Z}_{L+2,{3}} ({z},0,0,0)$. Essentially, this can be easily seen in the generating function, by just putting $f(0,0,0) = xf(1,1,1) = x^2f(2,2,2)$ (omitting $x$ inside the function argument as it is understood). Similarly, we can see that only those values of ($h_1,h_2,h_3$) are required where $h_i \geq 1$. This gives us that the independent triad(s) are $(h_1,h_2,h_3) = \{(0,0,0),(0,0,1),(0,1,1),(0,1,2),(0,0,2),(0,2,2)\}$ reducing the dimensionality from 10 to 6. This method also works for a general $k$-mer on a $k \times L$ strip. Essentially, the argument becomes that only the triad(s) where there is \textit{at least one value equal to 0} forms the independent basis. Thus, using these, we can form a transfer matrix formulation in the generating function formalism:
\begin{align}
    &MF = X, \\
    &F = M^{-1}X.
\end{align}
\begin{figure*}
    \centering
    
    \includegraphics[width=2\columnwidth]{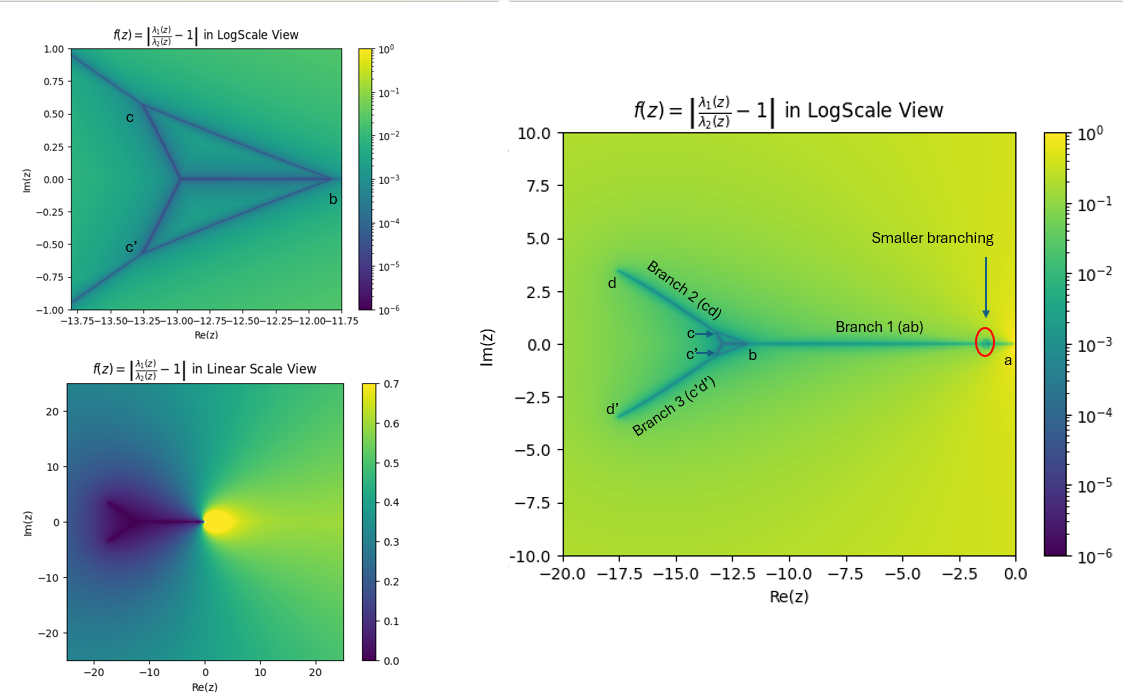}
    \caption{Heatmap images of the absolute ratio distance from 1. Image on the right highlights the eight (possibly curved) segments of {zeros} with the important endpoints ($a = -0.0716, b =-11.82, c = -13.25 + i0.25, d = -17.47 +i3.41$, $c',d'$ are complex conjugates of points $c$ and $d$ respectively). Top left heatmap gives a closer look on the branching starting from point $b$, while bottom left image is a zoomed-out linear scale view of the same scenario. The colorbar scale highlights $\delta$, the distance of the absolute value of ratio from 1.}
    \label{fig:heatmaps}
\end{figure*}
\begin{figure}
        \centering
        \includegraphics[width=\columnwidth]{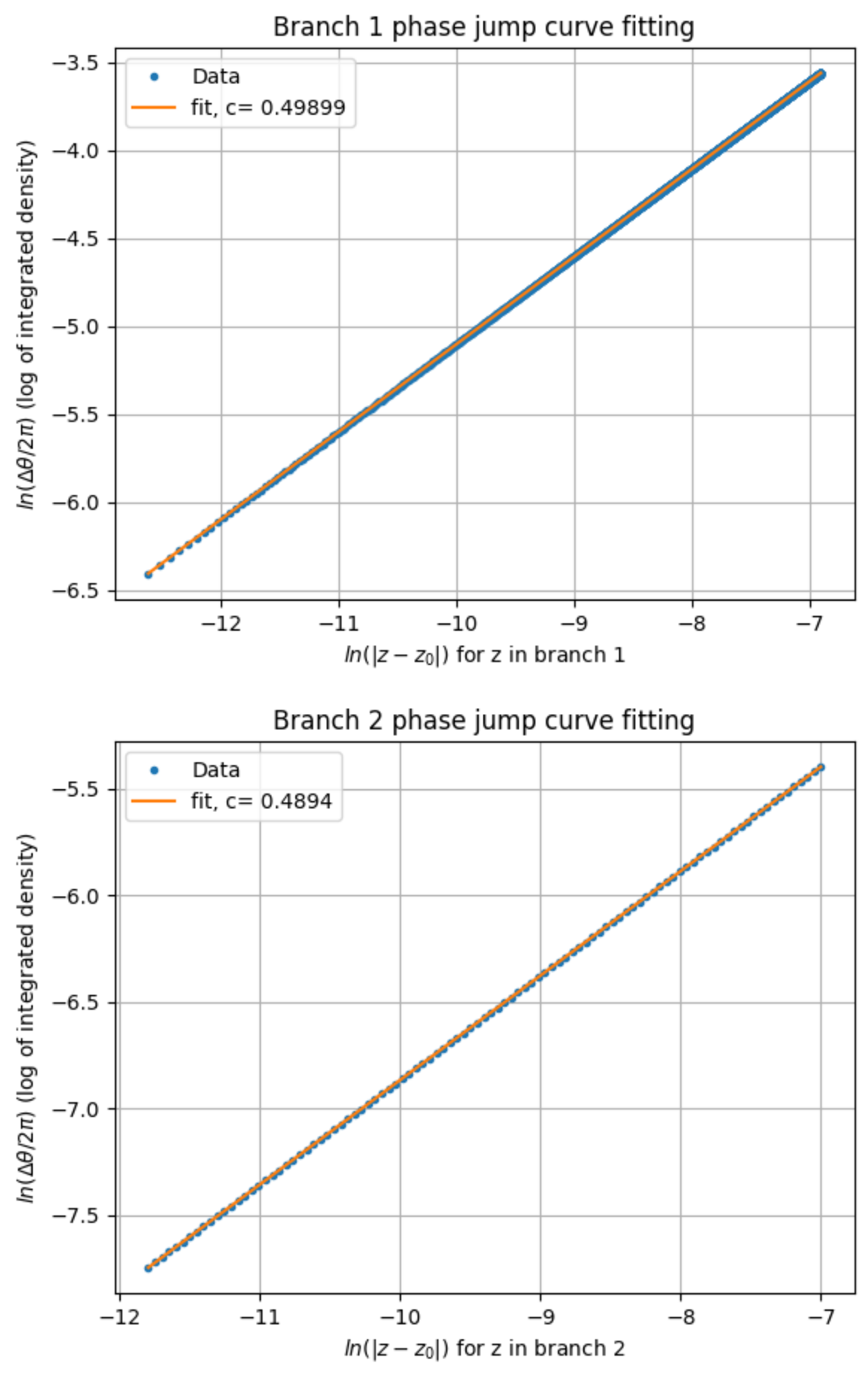}
        \caption{Log-Log graphs of integrated density {versus} distance from singularity points in branches 1 {(upper panel)} and 2 {(lower panel)}. The linear curve fit's slope essentially gives us the critical exponents for the density power-law relations very close to the singularity. We notice that both exponents are very close to $0.5$}
        \label{fig:branches}
\end{figure}
where $F$ is the column vector of generating functions, and $X$ is a constant vector.\\

We have also seen that our current transfer matrix method does not work very well for large $L$ ($L \approx 1000$ seems to be the current maximum if the system is allowed to run for 1 day). Numerical errors also creep in at higher values of $L$ which are difficult to control. Another method, which can be much faster, is listed: Consider the partition function ${Z}_L(z)$, and let $\{\lambda_i(z)\}, 1 \leq i \leq 10$, such that $|\lambda_i (z)| > |\lambda_j (z)|$ for $i > j$. It is easily seen that ${Z}_L(z) = \sum_i \lambda_i (z)^L$ (normalized such that any coefficients $c_i$ are absorbed into the eigenvalues). Let us only consider the first two largest eigenvalues, as in Sec.~\ref{sec:level2_3}, such that:
\begin{align}
    \tilde{{Z}}_{L,{3}}(z) = \lambda_1 (z)^L + \lambda_2 (z)^L.
\end{align}
Numerically, it is easy to determine the phase (values of $\theta$, analogous to the one as seen in Eq.~\eqref{eq:theta}) in the complex z-plane where $\frac{|\lambda_1 (z)|}{|\lambda_2 (z)|} = 1$. We just take any starting point $z_0$ and move in the direction which decreases (or increases, depending on the initial value) the value of $\frac{|\lambda_1 (z)|}{|\lambda_2 (z)|}$ the fastest, {very similar to the ideas in~\cite{janke2001,fisher1978,alves2000,glasser2001}}. Once we reach a value of $z$ where the ratio is $ \approx 1$ to an agreed accuracy, we find other such values of $z$ by moving in a direction where this condition is maintained. Once we have the line where the {zeros} will be in the large $L$ limit, as we move along the line, the density of {zeros} depends on how the value of $\theta$ fluctuates, which gives us essentially the integrated density of {zeros}. We analyze the jump in phase for $\ln(\lambda_{max}(z))$ across the branch cuts defined by the {zeros} of the partition function and relate it to the integrated density of {zeros} (see \appref{apn:app_2} for details). 

The results complement the finite $L$ numerics. \figref{fig:heatmaps}, highlights the heatmap images of this algorithm, where the colorbar indicates the absolute distance of the ratio from 1. \figref{fig:branches} highlights the natural log of integrated density (see \eqnref{eqn:int_density} in \appref{apn:app_2} for how integrated density is related to phase jump) {versus} natural log of distance from singularity plots on two of the three main branches (as branch 2 and 3 are complex conjugate, they have the same phase jump plots) considered on the heatmap, as a function of the increasing distance from singularity considered for the branch. We find that the curves are of the form $f(z) = a|z-z_0|^c$, $c_{num} = 0.49899$ for branch 1 with $z_0 = -0.07164738$, while $c_{num} = 0.4894$ for branch 2 with $z_0 = -17.465002999 + i3.410000022$. Note that for branch 1, point $a$ is equivalent to $z_0$, while for branch 2, it is point $d$ (see \figref{fig:heatmaps} right figure).

\section{Discussion and Outlook}
\label{sec:sec4}
{In summary, this study showcases the power of transfer matrix recursion methods in analyzing the partition function zeros of $k$-mers on finite lattices, leading to insights into the underlying critical behavior and universality classes. We validated our approach by first considering the well-understood case of dimers ($k = 2$) on both a one-dimensional line and a $2 \times L$ strip. Our findings confirmed that the partition function zeros lie exclusively on the negative real axis, with values ranging from $z_0 = -0.25$ up to $-\infty$ for the former, and $z_0 = -0.13826671$ up to $-\infty$ in the latter case (cf. \figref{fig:fig_2}), is consistent with theoretical predictions in previous works. This successful verification of our methodology provided a strong foundation for extending the study to higher-order $k$-mers.}

{Building on this, we investigated the case of trimers ($k = 3$) on $3 \times L$ strips, employing symmetry reductions to manage the computational complexity of our transfer matrix approach. Interestingly, unlike the dimer case where the partition function zeros remain unbounded, we observed that trimer partition function zeros are confined within the complex plane (cf. \figref{fig:heatmaps}). The presence of imaginary components in the zeros is not unexpected, as it naturally arises in constrained finite-size systems involving trimers. Our analysis further revealed that the density of zeros along both critical lines exhibits a critical exponent close to 0.5. This finding aligns with theoretical expectations that trimers on a 3 × L strip share the same universality class as dimers on a ladder system. The observed exponent equivalence at the endpoints of the partition function zeros suggests a deeper structural connection between these systems, reinforcing the idea that $k$-mer models on constrained lattices may exhibit universal scaling properties.}\\

{This work is designed to open up new avenues for the study of statistical mechanics of $k$-mers on finite and infinite lattices. Our current transfer matrix formalism can be readily extended to larger values of $k$, enabling the systematic exploration of $k$-mer models on general 2D lattices. A particularly intriguing direction is the numerical investigation of trimers (or even general $k$-mers) on generic $k \times L$ structures, where, unlike dimers on a 2D graph, theoretical results remain scarce. Understanding the behavior of $k$-mers in such settings may reveal novel universality classes and shed light on phase transitions in constrained interacting systems. Beyond the study of $k$-mers, our approach provides a robust computational framework that can be applied to various lattice-based statistical models, including those with hard-core exclusions, percolation, and adsorption processes. Additionally, investigating larger lattice sizes and extending the analysis to three-dimensional systems could offer new insights into the complex interplay between geometry and statistical mechanics. Integrating these numerical studies with field-theoretic approaches and renormalization group techniques may provide a deeper theoretical understanding of constrained lattice models and their broader implications for condensed matter physics.}


\begin{acknowledgments}
It is a pleasure to thank Prof. Deepak Dhar, {Prof. Elio K\"onig and Nisarg Chadha} for their valuable guidance and insights throughout this work. The author acknowledges funding from NIUS fellowship, TIFR Mumbai, and support from the KVPY program, DST India. 
All codes are made public on the GitHub link \url{https://github.com/lo568los/NIUS-Project}.\\
\end{acknowledgments}

\appendix

\section{Derivation of the transfer matrix}
\label{apn:app_1}

Here we will essentially sketch the derivation of the transfer matrix utilizing the first formulation. The key point to notice is that all $Z_i(y)$ for $y \leq L-3$ are covered amongst the configurations of $L-2\leq y \leq L + 1$ in $X_L$ (see \eqnref{eqn:tm_main}). To see this, we show in \figref{fig:Z2_config} the two contributions from the lowest strip length configuration which is $Z_2(L-2)$ for $Z_1(L+2)$ giving the $2z^2 Z_2(L-2)$ term:

\begin{figure}
    \centering
    
    \includegraphics[width=0.8\columnwidth]{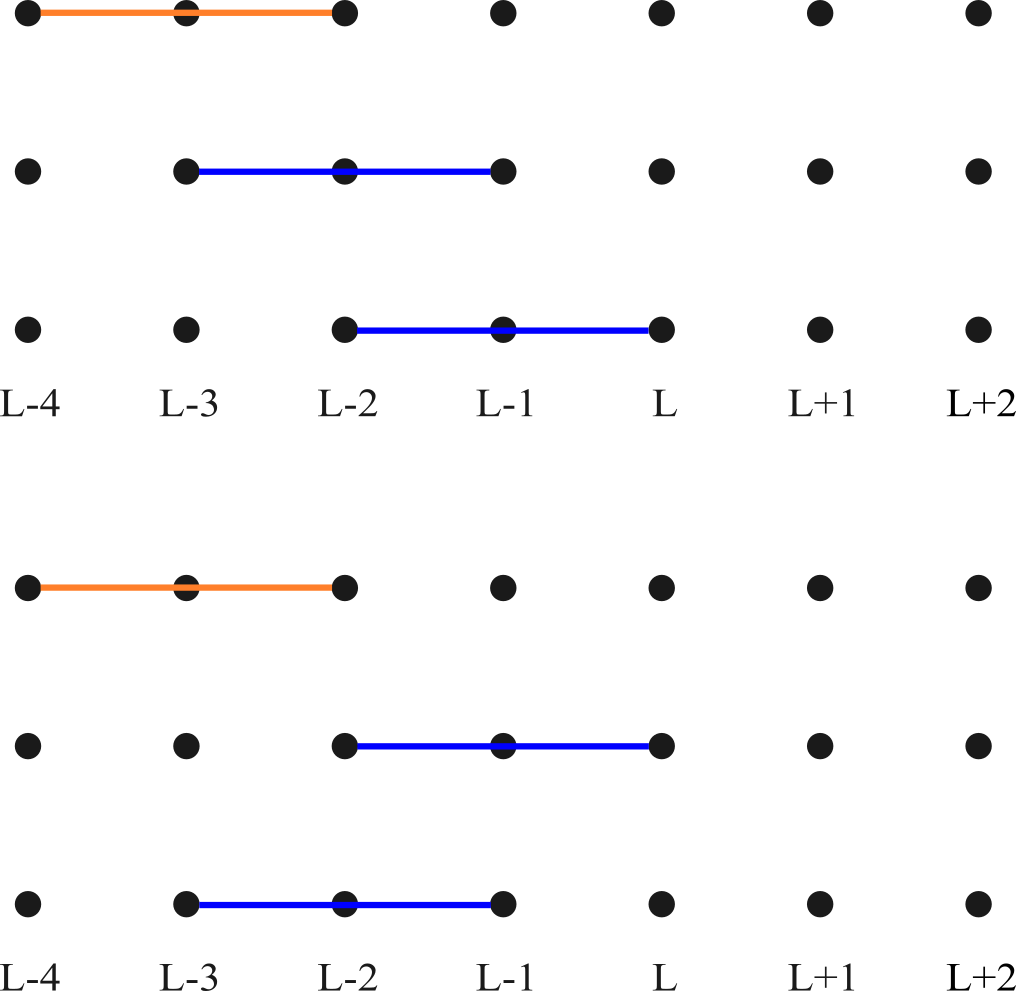}\\
    \caption{Contribution to $Z_1(L+2)$ from $Z_2(L-2)$ (orange) and two horizontal trimers (blue) giving the $z^2$ term in the product. Since for each $Z_2(L-2)$ configuration, there are two ways of placing the two trimers, hence the term $2z^2 Z_2(L-2)$.}
    \label{fig:Z2_config}
\end{figure}

No other $L-2$ term exists, and also no lower strip length term ($y < L-2$) is required in the recursion, as they have already been covered in the $L+1$ length terms. All the other terms of the recursion are also formed in the same fashion. $Z_1(y)$ for $y < L + 1$ is covered under $Z_1(L+1)$. Similarly $Z_4(y)$ for $y < L$ is also covered under $Z_1(L+1)$, while $Z_3(L-2)$ is covered by $Z_2(L), Z_2(L-1)$ and $Z_1(L+1)$.\\

The transformation from the first notation to the second one is also easily understood with pictorial representations. The key idea is that {the} sum of configurations {by permuting} some values of $h_1,h_2,h_3$ essentially equals {a} sum of configurations in the first notation. We trivially note down that $Z_1(L+2) = Z_{L,{3}}({z},0,0,0)$, $Z_4(L+1) = Z_{L,{3}}({z},1,1,1)$ and $Z_2(L+1) = Z_{L-1,{3}}({z},2,0,0) + Z_{L-1,{3}}({z},0,2,0) + Z_{L-1,{3}}({z},0,0,2)$. We {now} look at $h_1 = 0,~h_2 =0,~h_3 =1$ and its permutations, and see its sum, i.e. $Z_{L,{3}}({z},0,0,1) + Z_{L,{3}}({z},0,1,0) + Z_{L,{3}}({z},1,0,0) = 3Z_{L,{3}}({z},0,0,1)$ in \figref{fig:transform}. Essentially, we see that:

\begin{figure}
    \centering
    
    \includegraphics[width=\columnwidth]{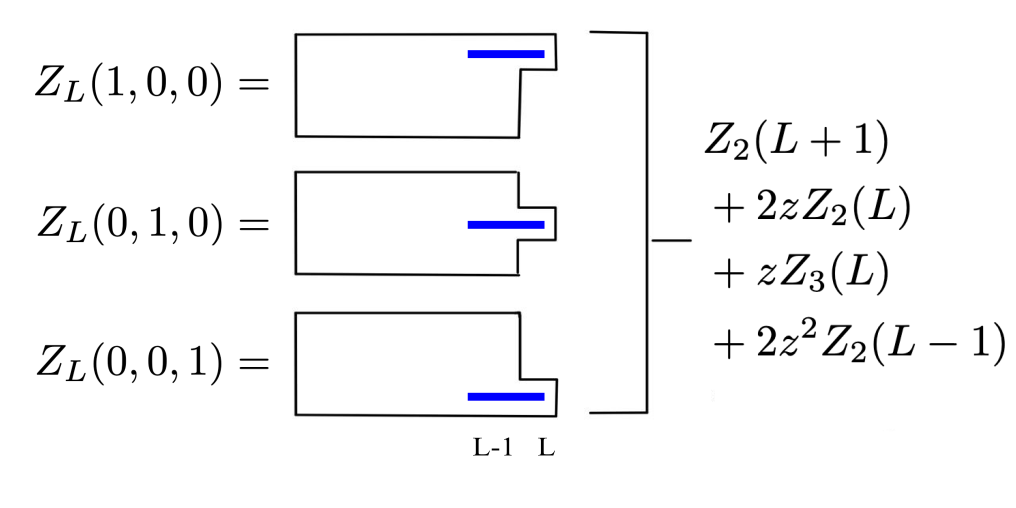}
    \caption{An example for showing {the} transformation from {the} first notation to the second notation, described in \secref{sec:level3_1}.}
    \label{fig:transform}
\end{figure}
\begin{align}
    3Z_{L,,{3}}({z},0,0,1) &= Z_2(L+1) + 2zZ_2(L) +zZ_3(L) \notag \\ 
    &+ 2z^2{Z_2(L-1)}.
\end{align}
The R.H.S contains all the configurations in the first notation which satisfy the boundary conditions of the L.H.S in the second condition. We can similarly do the same for all the independent configurations of the partition function in the second notation, and arrive at the equation $\tilde{X}_{L} = U(z) X_{L + 1}$, where:
\begin{equation}
    \begin{aligned}
     &U(z) = \begin{pmatrix}
          1  & 0 & 0 & 0 &  0 & 0 & 0 & 0 & 0 & 0 \\
          0 & 0 & 0 & 0 &  1/3 & 0 & 0 & 2z/3 & z/3 & 2z^2/3 \\
          0 & 0 & 0 & 0 &  0 & 1/3 & 0 & z^2/3 & 0 & 0 \\
          0 & 0 & 0 & 0 &  0 & 0 & 1 & 0 & 0 & 0 \\
          0 & 1/3 & 0 & 0 &  0 & 0 & 0 & 0 & 0 & 0 \\
          0 & 0 & 1/3 & 0 &  0 & 0 & 0 & 0 & 0 & 0 \\
          0 & 0 & 0 & 1/3 &  0 & 0 & 0 & 0 & 0 & 0 \\
          0 & 0 & 0 & 0 &  z/3 & 0 & 0 & z^2/3 & 0 & 0 \\
          0 & 0 & 0 & 0 &  0 & z^2/3 & 0 & 0 & 0 & 0 \\
          0 & 0 & 0 & 0 &  z/3 & 0 & 0 & 0 & 0 & 0 \\
         \end{pmatrix}, \\[10pt]
    &\tilde{X}_L = [Z_{L,{3}}({z},0,0,0),  Z_{L,{3}}({z},0,0,1),  Z_{L,{3}}({z},0,1,1),   \\ \nonumber  
    &Z_{L,{3}}({z},1,1,1),Z_{L,{3}}({z},0,0,2), Z_{L,{3}}({z},0,2,2),  Z_{L,{3}}({z},2,2,2),  \\ \nonumber  
    &Z_{L,{3}}({z},0,1,2),Z_{L,{3}}({z},1,2,2), Z_{L,{3}}({z},1,1,2)]^{T}. \\ \nonumber
    \end{aligned}
\end{equation}
\section{Branch cuts for the phase jump plotting}
\label{apn:app_2}

Here we discuss the relation between the density of {zeros} per unit length, which is $\lim_{L \to \infty} \phi_L/L$, where $\phi_L$ is the density of {zeros} on a branch. We define $\phi_L^i(z)$ to be the density corresponding to branch $i = 1,2,3$ as according to \figref{fig:heatmaps}. Note that this density is different from $\rho(z)$, which is the density of filled sites per total number of sites in the lattice as a function of {fugacity}. We note the following expressions:
\begin{align}
    \rho(z) = \frac{z}{L}\frac{\partial \ln({Z}_L(z))}{\partial z},
\end{align}
where ${Z}_L(z)$ is the partition function for a finite strip of length $L$. We can write ${Z}_L(z) \approx \lambda_{max}^L(z)$, where $\lambda_{max}$ is the eigenvalue of the transfer matrix with the highest magnitude at $z$, and thus have:

\begin{figure}

    \centering
    
    \includegraphics[width=\columnwidth]{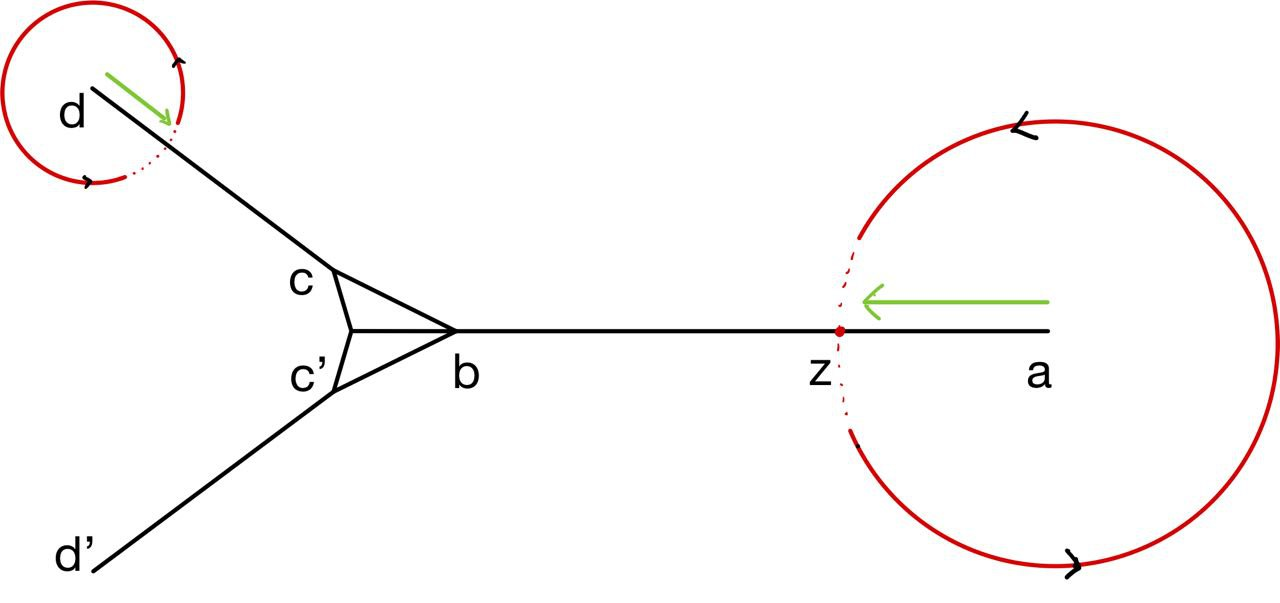}
    \caption{Branch cuts with examples of contour lines. In this scenario, points $a$ and $d$ are the singularity points ($z_0$) for branches 1 and 2, respectively. The green lines highlight the {zeros} encompassed by the contour lines (drawn in red)}
    \label{fig:contours}
\end{figure}

\begin{figure}
\hspace*{-5mm}
    \centering
    \includegraphics[width=\columnwidth]{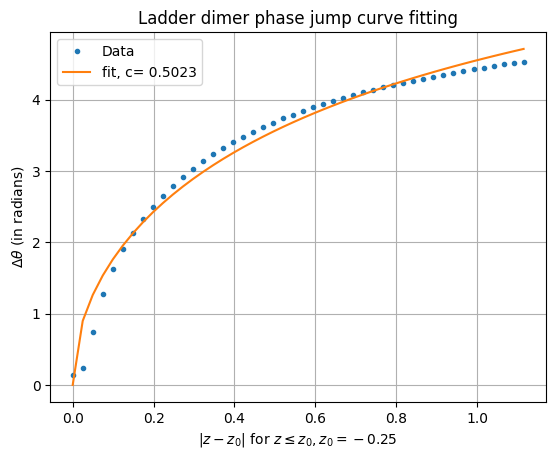}
    \caption{Phase jump plot with curve fitting of the configuration in \secref{sec:level2_2}. The fit value is for the integrated density of {zeros} curve}
    \label{fig:dimerphase}
\end{figure}

Define $\theta$ as the phase of the complex-valued function $\ln(\lambda_{max}(z - z_0))$, where $z_0$ is the endpoint of a branch cut, thus making it dependent on $z$ itself. Since the natural logarithm function is multivalued in the complex plane, we carefully choose the direction of our branch cuts (originating from the {zeros} of the partition function) such that the cuts overlap on the direction of the line of {zeros} and thus cancel out outside of these lines. \figref{fig:contours} highlights these branch cuts. These endpoints $z_0$ are chosen carefully such that when a contour integral is performed as in \figref{fig:contours}, the encompassed {zeros} are proportional to the phase jump across the branch cut discontinuity.  We formally note that if ${Z}_L(z) = |{Z}_L(z)|e^{i\theta}$:
\begin{align}
    \frac{\ln({Z}_L(z))}{L} = \frac{\ln(|{Z}_L(z)|)}{L} + i\frac{\theta}{L}.
\end{align}

Here, the $\theta/L$ term is multivalued as $\theta \to \theta + 2\pi n$ is also a valid phase. If we take each zero to have an analogous charge of $1/L$, crossing the branch cut starting from one such "charge" will give a phase jump of $2\pi/L$. Similarly, if the contour crosses the common branch cut from two charges, the phase jump will be $4\pi/L$. Generalizing this, the phase jump ($\Delta\theta$) is equal to $2\pi n/L$, where $n$ is the number of encompassed {zeros}. But note that in the thermodynamic limit, $\lim_{L \to \infty} n/L = \int_{z_0}^z \phi (t)dt$, where $z$ is the point encompassing the $n$ {zeros}.
\begin{align}
    \label{eqn:int_density}
    2\pi \int_{z_0}^z \phi (t)dt &= \Delta \theta, \\ \nonumber
    \int_{z_0}^z \phi (t)dt &= \frac{\Delta \theta}{2\pi} .
\end{align}
We can extract the power-law expressions of the density of {zeros} of the form of $|z-z_0|^c$ close to the singularity $z_0$ (as expressed in the original Lee-Yang paper), the integrated density of {zeros} will be of the form of $|z-z_0|^{c+1}$. We can find the value of $c$ by curve-fitting power law expressions to our phase jump plots. {This idea has also been presented in~\cite{itzykson1983,kenna2004,bena2005,kortman1971}}. As a test of accuracy for this method, \figref{fig:dimerphase} shows the curve fitting to the phase jump plot on the dimer on ladder problem (refer \secref{sec:level2_2}), for which it is known that $c = -0.5$. With our curve fit of the form $f(x) = ax^c$, we obtained $c_{num} = 0.5023$, giving the power law exponent as equal to $-0.4977$, which is reasonably close to the correct theoretical value.

\nocite{*}
\bibliography{aipsamp}

\providecommand{\noopsort}[1]{}\providecommand{\singleletter}[1]{#1}%
\begin{thebibliography}{44}%
\makeatletter
\providecommand \@ifxundefined [1]{%
 \@ifx{#1\undefined}
}%
\providecommand \@ifnum [1]{%
 \ifnum #1\expandafter \@firstoftwo
 \else \expandafter \@secondoftwo
 \fi
}%
\providecommand \@ifx [1]{%
 \ifx #1\expandafter \@firstoftwo
 \else \expandafter \@secondoftwo
 \fi
}%
\providecommand \natexlab [1]{#1}%
\providecommand \enquote  [1]{``#1''}%
\providecommand \bibnamefont  [1]{#1}%
\providecommand \bibfnamefont [1]{#1}%
\providecommand \citenamefont [1]{#1}%
\providecommand \href@noop [0]{\@secondoftwo}%
\providecommand \href [0]{\begingroup \@sanitize@url \@href}%
\providecommand \@href[1]{\@@startlink{#1}\@@href}%
\providecommand \@@href[1]{\endgroup#1\@@endlink}%
\providecommand \@sanitize@url [0]{\catcode `\\12\catcode `\$12\catcode `\&12\catcode `\#12\catcode `\^12\catcode `\_12\catcode `\%12\relax}%
\providecommand \@@startlink[1]{}%
\providecommand \@@endlink[0]{}%
\providecommand \url  [0]{\begingroup\@sanitize@url \@url }%
\providecommand \@url [1]{\endgroup\@href {#1}{\urlprefix }}%
\providecommand \urlprefix  [0]{URL }%
\providecommand \Eprint [0]{\href }%
\providecommand \doibase [0]{https://doi.org/}%
\providecommand \selectlanguage [0]{\@gobble}%
\providecommand \bibinfo  [0]{\@secondoftwo}%
\providecommand \bibfield  [0]{\@secondoftwo}%
\providecommand \translation [1]{[#1]}%
\providecommand \BibitemOpen [0]{}%
\providecommand \bibitemStop [0]{}%
\providecommand \bibitemNoStop [0]{.\EOS\space}%
\providecommand \EOS [0]{\spacefactor3000\relax}%
\providecommand \BibitemShut  [1]{\csname bibitem#1\endcsname}%
\let\auto@bib@innerbib\@empty
\bibitem [{\citenamefont {van~der Waals}(1873)}]{vanderwaals1873}%
  \BibitemOpen
  \bibfield  {author} {\bibinfo {author} {\bibfnamefont {J.~D.}\ \bibnamefont {van~der Waals}},\ }\href@noop {} {\emph {\bibinfo {title} {Over de Continuïteit van den Gas- en Vloeistoftoestand}}}\ (\bibinfo  {publisher} {PhD Thesis, Leiden University},\ \bibinfo {year} {1873})\BibitemShut {NoStop}%
\bibitem [{\citenamefont {Maxwell}(1875)}]{maxwell1875}%
  \BibitemOpen
  \bibfield  {author} {\bibinfo {author} {\bibfnamefont {J.~C.}\ \bibnamefont {Maxwell}},\ }\bibfield  {title} {\bibinfo {title} {On the dynamical evidence of the molecular constitution of bodies},\ }\href@noop {} {\bibfield  {journal} {\bibinfo  {journal} {Nature}\ }\textbf {\bibinfo {volume} {11}},\ \bibinfo {pages} {357} (\bibinfo {year} {1875})}\BibitemShut {NoStop}%
\bibitem [{\citenamefont {Ehrenfest}(1933)}]{ehrenfest1933}%
  \BibitemOpen
  \bibfield  {author} {\bibinfo {author} {\bibfnamefont {P.}~\bibnamefont {Ehrenfest}},\ }\bibfield  {title} {\bibinfo {title} {Phase transitions in statistical mechanics},\ }\href@noop {} {\bibfield  {journal} {\bibinfo  {journal} {Proceedings of the Royal Netherlands Academy of Arts and Sciences}\ }\textbf {\bibinfo {volume} {36}},\ \bibinfo {pages} {153} (\bibinfo {year} {1933})}\BibitemShut {NoStop}%
\bibitem [{\citenamefont {Onsager}(1944)}]{onsager1944}%
  \BibitemOpen
  \bibfield  {author} {\bibinfo {author} {\bibfnamefont {L.}~\bibnamefont {Onsager}},\ }\bibfield  {title} {\bibinfo {title} {Crystal statistics. i. a two-dimensional model with an order-disorder transition},\ }\href@noop {} {\bibfield  {journal} {\bibinfo  {journal} {Physical Review}\ }\textbf {\bibinfo {volume} {65}},\ \bibinfo {pages} {117} (\bibinfo {year} {1944})}\BibitemShut {NoStop}%
\bibitem [{\citenamefont {Kadanoff}(1966)}]{kadanoff1966}%
  \BibitemOpen
  \bibfield  {author} {\bibinfo {author} {\bibfnamefont {L.~P.}\ \bibnamefont {Kadanoff}},\ }\bibfield  {title} {\bibinfo {title} {Scaling laws for ising models near ${T}_c$},\ }\href@noop {} {\bibfield  {journal} {\bibinfo  {journal} {Physics Physique Physics}\ }\textbf {\bibinfo {volume} {2}},\ \bibinfo {pages} {263} (\bibinfo {year} {1966})}\BibitemShut {NoStop}%
\bibitem [{\citenamefont {Wilson}(1971)}]{wilson1971}%
  \BibitemOpen
  \bibfield  {author} {\bibinfo {author} {\bibfnamefont {K.~G.}\ \bibnamefont {Wilson}},\ }\bibfield  {title} {\bibinfo {title} {Renormalization group and critical phenomena. i. renormalization group and the kadanoff scaling picture},\ }\href@noop {} {\bibfield  {journal} {\bibinfo  {journal} {Physical Review B}\ }\textbf {\bibinfo {volume} {4}},\ \bibinfo {pages} {3174} (\bibinfo {year} {1971})}\BibitemShut {NoStop}%
\bibitem [{\citenamefont {Yang}\ and\ \citenamefont {Lee}(1952)}]{leeyang1}%
  \BibitemOpen
  \bibfield  {author} {\bibinfo {author} {\bibfnamefont {C.~N.}\ \bibnamefont {Yang}}\ and\ \bibinfo {author} {\bibfnamefont {T.~D.}\ \bibnamefont {Lee}},\ }\bibfield  {title} {\bibinfo {title} {Statistical theory of equations of state and phase transitions. i. theory of condensation},\ }\href {https://doi.org/10.1103/PhysRev.87.404} {\bibfield  {journal} {\bibinfo  {journal} {Phys. Rev.}\ }\textbf {\bibinfo {volume} {87}},\ \bibinfo {pages} {404} (\bibinfo {year} {1952})}\BibitemShut {NoStop}%
\bibitem [{\citenamefont {Lee}\ and\ \citenamefont {Yang}(1952)}]{leeyang2}%
  \BibitemOpen
  \bibfield  {author} {\bibinfo {author} {\bibfnamefont {T.~D.}\ \bibnamefont {Lee}}\ and\ \bibinfo {author} {\bibfnamefont {C.~N.}\ \bibnamefont {Yang}},\ }\bibfield  {title} {\bibinfo {title} {Statistical theory of equations of state and phase transitions. ii. lattice gas and ising model},\ }\href {https://doi.org/10.1103/PhysRev.87.410} {\bibfield  {journal} {\bibinfo  {journal} {Phys. Rev.}\ }\textbf {\bibinfo {volume} {87}},\ \bibinfo {pages} {410} (\bibinfo {year} {1952})}\BibitemShut {NoStop}%
\bibitem [{\citenamefont {Fisher}(1965)}]{fisher}%
  \BibitemOpen
  \bibfield  {author} {\bibinfo {author} {\bibfnamefont {M.~E.}\ \bibnamefont {Fisher}},\ }\bibfield  {title} {\bibinfo {title} {The nature of critical points},\ }\href@noop {} {\bibfield  {journal} {\bibinfo  {journal} {Lectures in Theoretical Physics}\ }\textbf {\bibinfo {volume} {7C}},\ \bibinfo {pages} {1} (\bibinfo {year} {1965})}\BibitemShut {NoStop}%
\bibitem [{\citenamefont {Griffiths}(1969)}]{griffiths1969}%
  \BibitemOpen
  \bibfield  {author} {\bibinfo {author} {\bibfnamefont {R.~B.}\ \bibnamefont {Griffiths}},\ }\bibfield  {title} {\bibinfo {title} {Nonanalytic behavior above the critical point in a random ising ferromagnet},\ }\href@noop {} {\bibfield  {journal} {\bibinfo  {journal} {Physical Review Letters}\ }\textbf {\bibinfo {volume} {23}},\ \bibinfo {pages} {17} (\bibinfo {year} {1969})}\BibitemShut {NoStop}%
\bibitem [{\citenamefont {Temperley}\ and\ \citenamefont {Lieb}(1971)}]{temperley}%
  \BibitemOpen
  \bibfield  {author} {\bibinfo {author} {\bibfnamefont {H.~N.~V.}\ \bibnamefont {Temperley}}\ and\ \bibinfo {author} {\bibfnamefont {E.~H.}\ \bibnamefont {Lieb}},\ }\bibfield  {title} {\bibinfo {title} {Relations between the ‘percolation’ and ‘colouring’ problem and other graph-theoretical problems associated with regular planar lattices: Some exact results for the ‘percolation’ problem},\ }\href@noop {} {\bibfield  {journal} {\bibinfo  {journal} {Proceedings of the Royal Society of London. Series A, Mathematical and Physical Sciences}\ }\textbf {\bibinfo {volume} {322}},\ \bibinfo {pages} {251} (\bibinfo {year} {1971})}\BibitemShut {NoStop}%
\bibitem [{\citenamefont {Fisher}(1978{\natexlab{a}})}]{fisher3}%
  \BibitemOpen
  \bibfield  {author} {\bibinfo {author} {\bibfnamefont {M.~E.}\ \bibnamefont {Fisher}},\ }\bibfield  {title} {\bibinfo {title} {Yang-lee edge singularity and ${\ensuremath{\phi}}^{3}$ field theory},\ }\href {https://doi.org/10.1103/PhysRevLett.40.1610} {\bibfield  {journal} {\bibinfo  {journal} {Phys. Rev. Lett.}\ }\textbf {\bibinfo {volume} {40}},\ \bibinfo {pages} {1610} (\bibinfo {year} {1978}{\natexlab{a}})}\BibitemShut {NoStop}%
\bibitem [{\citenamefont {Baxter}()}]{baxter}%
  \BibitemOpen
  \bibfield  {author} {\bibinfo {author} {\bibfnamefont {R.~J.}\ \bibnamefont {Baxter}},\ }\bibinfo {title} {Exactly solved models in statistical mechanics},\ in\ \href {https://doi.org/10.1142/9789814415255_0002} {\emph {\bibinfo {booktitle} {Integrable Systems in Statistical Mechanics}}},\ pp.\ \bibinfo {pages} {5--63}\BibitemShut {NoStop}%
\bibitem [{\citenamefont {Assis}\ \emph {et~al.}(2013)\citenamefont {Assis}, \citenamefont {Jacobsen}, \citenamefont {Jensen}, \citenamefont {Maillard},\ and\ \citenamefont {McCoy}}]{Assis_Jacobsen_Jensen_Maillard_McCoy_2013}%
  \BibitemOpen
  \bibfield  {author} {\bibinfo {author} {\bibfnamefont {M.}~\bibnamefont {Assis}}, \bibinfo {author} {\bibfnamefont {J.~L.}\ \bibnamefont {Jacobsen}}, \bibinfo {author} {\bibfnamefont {I.}~\bibnamefont {Jensen}}, \bibinfo {author} {\bibfnamefont {J.-M.}\ \bibnamefont {Maillard}},\ and\ \bibinfo {author} {\bibfnamefont {B.~M.}\ \bibnamefont {McCoy}},\ }\bibfield  {title} {\bibinfo {title} {The hard hexagon partition function for complex fugacity},\ }\href {https://doi.org/10.1088/1751-8113/46/44/445202} {\bibfield  {journal} {\bibinfo  {journal} {Journal of Physics A: Mathematical and Theoretical}\ }\textbf {\bibinfo {volume} {46}},\ \bibinfo {pages} {445202} (\bibinfo {year} {2013})}\BibitemShut {NoStop}%
\bibitem [{\citenamefont {de~Gennes}\ and\ \citenamefont {Prost}(1993)}]{degennes1993}%
  \BibitemOpen
  \bibfield  {author} {\bibinfo {author} {\bibfnamefont {P.~G.}\ \bibnamefont {de~Gennes}}\ and\ \bibinfo {author} {\bibfnamefont {J.}~\bibnamefont {Prost}},\ }\href@noop {} {\emph {\bibinfo {title} {The Physics of Liquid Crystals}}}\ (\bibinfo  {publisher} {Oxford University Press},\ \bibinfo {year} {1993})\BibitemShut {NoStop}%
\bibitem [{\citenamefont {Flory}(1953)}]{flory1953}%
  \BibitemOpen
  \bibfield  {author} {\bibinfo {author} {\bibfnamefont {P.~J.}\ \bibnamefont {Flory}},\ }\href@noop {} {\emph {\bibinfo {title} {Principles of Polymer Chemistry}}}\ (\bibinfo  {publisher} {Cornell University Press},\ \bibinfo {year} {1953})\BibitemShut {NoStop}%
\bibitem [{\citenamefont {Whitesides}\ and\ \citenamefont {Grzybowski}(2002)}]{whitesides2002}%
  \BibitemOpen
  \bibfield  {author} {\bibinfo {author} {\bibfnamefont {G.~M.}\ \bibnamefont {Whitesides}}\ and\ \bibinfo {author} {\bibfnamefont {B.}~\bibnamefont {Grzybowski}},\ }\bibfield  {title} {\bibinfo {title} {Self-assembly at all scales},\ }\href@noop {} {\bibfield  {journal} {\bibinfo  {journal} {Science}\ }\textbf {\bibinfo {volume} {295}},\ \bibinfo {pages} {2418} (\bibinfo {year} {2002})}\BibitemShut {NoStop}%
\bibitem [{\citenamefont {Fraden}\ \emph {et~al.}(1989)\citenamefont {Fraden}, \citenamefont {Maret}, \citenamefont {Caspar},\ and\ \citenamefont {Meyer}}]{tobacomosaic}%
  \BibitemOpen
  \bibfield  {author} {\bibinfo {author} {\bibfnamefont {S.}~\bibnamefont {Fraden}}, \bibinfo {author} {\bibfnamefont {G.}~\bibnamefont {Maret}}, \bibinfo {author} {\bibfnamefont {D.~L.~D.}\ \bibnamefont {Caspar}},\ and\ \bibinfo {author} {\bibfnamefont {R.~B.}\ \bibnamefont {Meyer}},\ }\bibfield  {title} {\bibinfo {title} {Isotropic-nematic phase transition and angular correlations in isotropic suspensions of tobacco mosaic virus},\ }\href {https://doi.org/10.1103/PhysRevLett.63.2068} {\bibfield  {journal} {\bibinfo  {journal} {Phys. Rev. Lett.}\ }\textbf {\bibinfo {volume} {63}},\ \bibinfo {pages} {2068} (\bibinfo {year} {1989})}\BibitemShut {NoStop}%
\bibitem [{\citenamefont {Mogilner}\ and\ \citenamefont {Oster}(2003)}]{mogilner2003}%
  \BibitemOpen
  \bibfield  {author} {\bibinfo {author} {\bibfnamefont {A.}~\bibnamefont {Mogilner}}\ and\ \bibinfo {author} {\bibfnamefont {G.}~\bibnamefont {Oster}},\ }\bibfield  {title} {\bibinfo {title} {Polymer motors: pushing out the front and pulling up the back},\ }\href@noop {} {\bibfield  {journal} {\bibinfo  {journal} {Current Biology}\ }\textbf {\bibinfo {volume} {13}},\ \bibinfo {pages} {R721} (\bibinfo {year} {2003})}\BibitemShut {NoStop}%
\bibitem [{\citenamefont {Islam}\ \emph {et~al.}(2004)\citenamefont {Islam}, \citenamefont {Alsayed}, \citenamefont {Dogic}, \citenamefont {Zhang}, \citenamefont {Lubensky},\ and\ \citenamefont {Yodh}}]{nanotube}%
  \BibitemOpen
  \bibfield  {author} {\bibinfo {author} {\bibfnamefont {M.~F.}\ \bibnamefont {Islam}}, \bibinfo {author} {\bibfnamefont {A.~M.}\ \bibnamefont {Alsayed}}, \bibinfo {author} {\bibfnamefont {Z.}~\bibnamefont {Dogic}}, \bibinfo {author} {\bibfnamefont {J.}~\bibnamefont {Zhang}}, \bibinfo {author} {\bibfnamefont {T.~C.}\ \bibnamefont {Lubensky}},\ and\ \bibinfo {author} {\bibfnamefont {A.~G.}\ \bibnamefont {Yodh}},\ }\bibfield  {title} {\bibinfo {title} {Nematic nanotube gels},\ }\href {https://doi.org/10.1103/PhysRevLett.92.088303} {\bibfield  {journal} {\bibinfo  {journal} {Phys. Rev. Lett.}\ }\textbf {\bibinfo {volume} {92}},\ \bibinfo {pages} {088303} (\bibinfo {year} {2004})}\BibitemShut {NoStop}%
\bibitem [{\citenamefont {Livolant}\ and\ \citenamefont {Leforestier}(1996)}]{livolant1996}%
  \BibitemOpen
  \bibfield  {author} {\bibinfo {author} {\bibfnamefont {F.}~\bibnamefont {Livolant}}\ and\ \bibinfo {author} {\bibfnamefont {A.}~\bibnamefont {Leforestier}},\ }\bibfield  {title} {\bibinfo {title} {Condensed phases of dna: Structures and phase transitions},\ }\href@noop {} {\bibfield  {journal} {\bibinfo  {journal} {Progress in Polymer Science}\ }\textbf {\bibinfo {volume} {21}},\ \bibinfo {pages} {1115} (\bibinfo {year} {1996})}\BibitemShut {NoStop}%
\bibitem [{\citenamefont {Kasteleyn}(1961)}]{kasteleyn1961}%
  \BibitemOpen
  \bibfield  {author} {\bibinfo {author} {\bibfnamefont {P.~W.}\ \bibnamefont {Kasteleyn}},\ }\bibfield  {title} {\bibinfo {title} {The statistics of dimers on a lattice},\ }\href@noop {} {\bibfield  {journal} {\bibinfo  {journal} {Physica}\ }\textbf {\bibinfo {volume} {27}},\ \bibinfo {pages} {1209} (\bibinfo {year} {1961})}\BibitemShut {NoStop}%
\bibitem [{\citenamefont {Frenkel}\ and\ \citenamefont {Eppenga}(1985)}]{frenkel1985}%
  \BibitemOpen
  \bibfield  {author} {\bibinfo {author} {\bibfnamefont {D.}~\bibnamefont {Frenkel}}\ and\ \bibinfo {author} {\bibfnamefont {R.}~\bibnamefont {Eppenga}},\ }\bibfield  {title} {\bibinfo {title} {Evidence for algebraic orientational order in a two-dimensional hard-core nematic},\ }\href@noop {} {\bibfield  {journal} {\bibinfo  {journal} {Physical Review A}\ }\textbf {\bibinfo {volume} {31}},\ \bibinfo {pages} {1776} (\bibinfo {year} {1985})}\BibitemShut {NoStop}%
\bibitem [{\citenamefont {McCoy}(2010)}]{mccoy2010}%
  \BibitemOpen
  \bibfield  {author} {\bibinfo {author} {\bibfnamefont {B.~M.}\ \bibnamefont {McCoy}},\ }\href@noop {} {\emph {\bibinfo {title} {Advanced Statistical Mechanics}}}\ (\bibinfo  {publisher} {Oxford University Press},\ \bibinfo {year} {2010})\BibitemShut {NoStop}%
\bibitem [{\citenamefont {Nisbet}\ and\ \citenamefont {Melkonyan}(2015)}]{nisbet2015}%
  \BibitemOpen
  \bibfield  {author} {\bibinfo {author} {\bibfnamefont {D.}~\bibnamefont {Nisbet}}\ and\ \bibinfo {author} {\bibfnamefont {G.~Y.}\ \bibnamefont {Melkonyan}},\ }\bibfield  {title} {\bibinfo {title} {Phase behavior of interacting k-mers and its implications in statistical mechanics},\ }\href@noop {} {\bibfield  {journal} {\bibinfo  {journal} {Journal of Statistical Mechanics: Theory and Experiment}\ }\textbf {\bibinfo {volume} {2015}},\ \bibinfo {pages} {P03007} (\bibinfo {year} {2015})}\BibitemShut {NoStop}%
\bibitem [{\citenamefont {Heilmann}\ and\ \citenamefont {Lieb}(1972)}]{Heilmann_Lieb_1972}%
  \BibitemOpen
  \bibfield  {author} {\bibinfo {author} {\bibfnamefont {O.~J.}\ \bibnamefont {Heilmann}}\ and\ \bibinfo {author} {\bibfnamefont {E.~H.}\ \bibnamefont {Lieb}},\ }\bibfield  {title} {\bibinfo {title} {Theory of monomer-dimer systems},\ }\href {https://doi.org/10.1007/bf01877590} {\bibfield  {journal} {\bibinfo  {journal} {Communications in Mathematical Physics}\ }\textbf {\bibinfo {volume} {25}},\ \bibinfo {pages} {190–232} (\bibinfo {year} {1972})}\BibitemShut {NoStop}%
\bibitem [{\citenamefont {Huang}(1987)}]{huang1987}%
  \BibitemOpen
  \bibfield  {author} {\bibinfo {author} {\bibfnamefont {K.}~\bibnamefont {Huang}},\ }\href@noop {} {\emph {\bibinfo {title} {Statistical Mechanics}}},\ \bibinfo {edition} {2nd}\ ed.\ (\bibinfo  {publisher} {John Wiley \& Sons},\ \bibinfo {year} {1987})\BibitemShut {NoStop}%
\bibitem [{\citenamefont {Landau}\ and\ \citenamefont {Lifshitz}(1980)}]{landau1980}%
  \BibitemOpen
  \bibfield  {author} {\bibinfo {author} {\bibfnamefont {L.~D.}\ \bibnamefont {Landau}}\ and\ \bibinfo {author} {\bibfnamefont {E.~M.}\ \bibnamefont {Lifshitz}},\ }\href@noop {} {\emph {\bibinfo {title} {Statistical Physics, Part 1}}},\ \bibinfo {edition} {3rd}\ ed.,\ \bibinfo {series} {Course of Theoretical Physics}, Vol.~\bibinfo {volume} {5}\ (\bibinfo  {publisher} {Pergamon Press},\ \bibinfo {year} {1980})\BibitemShut {NoStop}%
\bibitem [{\citenamefont {Pathria}\ and\ \citenamefont {Beale}(2011)}]{pathria2011}%
  \BibitemOpen
  \bibfield  {author} {\bibinfo {author} {\bibfnamefont {R.~K.}\ \bibnamefont {Pathria}}\ and\ \bibinfo {author} {\bibfnamefont {P.~D.}\ \bibnamefont {Beale}},\ }\href@noop {} {\emph {\bibinfo {title} {Statistical Mechanics}}},\ \bibinfo {edition} {3rd}\ ed.\ (\bibinfo  {publisher} {Elsevier},\ \bibinfo {year} {2011})\BibitemShut {NoStop}%
\bibitem [{\citenamefont {Fowler}\ and\ \citenamefont {Guggenheim}(1939)}]{fowler1939}%
  \BibitemOpen
  \bibfield  {author} {\bibinfo {author} {\bibfnamefont {R.~H.}\ \bibnamefont {Fowler}}\ and\ \bibinfo {author} {\bibfnamefont {E.~A.}\ \bibnamefont {Guggenheim}},\ }\bibfield  {title} {\bibinfo {title} {Statistical thermodynamics},\ }\href@noop {} {\bibfield  {journal} {\bibinfo  {journal} {Cambridge University Press}\ } (\bibinfo {year} {1939})}\BibitemShut {NoStop}%
\bibitem [{\citenamefont {Yeomans}(1992)}]{yeomans1992}%
  \BibitemOpen
  \bibfield  {author} {\bibinfo {author} {\bibfnamefont {J.~M.}\ \bibnamefont {Yeomans}},\ }\href@noop {} {\emph {\bibinfo {title} {Statistical Mechanics of Phase Transitions}}}\ (\bibinfo  {publisher} {Oxford University Press},\ \bibinfo {year} {1992})\BibitemShut {NoStop}%
\bibitem [{\citenamefont {Hill}(1962)}]{hill1962}%
  \BibitemOpen
  \bibfield  {author} {\bibinfo {author} {\bibfnamefont {T.~L.}\ \bibnamefont {Hill}},\ }\bibfield  {title} {\bibinfo {title} {Fugacity and activity in lattice gases and solutions},\ }\href {https://doi.org/10.1063/1.1732552} {\bibfield  {journal} {\bibinfo  {journal} {The Journal of Chemical Physics}\ }\textbf {\bibinfo {volume} {36}},\ \bibinfo {pages} {318} (\bibinfo {year} {1962})}\BibitemShut {NoStop}%
\bibitem [{\citenamefont {George Brown~Arfken}(2011)}]{Arfken2011}%
  \BibitemOpen
  \bibfield  {author} {\bibinfo {author} {\bibfnamefont {F.~E.~H.}\ \bibnamefont {George Brown~Arfken}, \bibfnamefont {Hans Jürgen~Weber}},\ }\href@noop {} {\bibinfo {title} {Mathematical methods for physicists}} (\bibinfo {year} {2011})\BibitemShut {NoStop}%
\bibitem [{\citenamefont {Janke}\ and\ \citenamefont {Kenna}(2001)}]{janke2001}%
  \BibitemOpen
  \bibfield  {author} {\bibinfo {author} {\bibfnamefont {W.}~\bibnamefont {Janke}}\ and\ \bibinfo {author} {\bibfnamefont {R.}~\bibnamefont {Kenna}},\ }\bibfield  {title} {\bibinfo {title} {Phase transition strengths from the density of partition function zeros},\ }\href {https://doi.org/10.1023/A:1004871610170} {\bibfield  {journal} {\bibinfo  {journal} {Journal of Statistical Physics}\ }\textbf {\bibinfo {volume} {102}},\ \bibinfo {pages} {1211–1227} (\bibinfo {year} {2001})}\BibitemShut {NoStop}%
\bibitem [{\citenamefont {Fisher}(1978{\natexlab{b}})}]{fisher1978}%
  \BibitemOpen
  \bibfield  {author} {\bibinfo {author} {\bibfnamefont {M.~E.}\ \bibnamefont {Fisher}},\ }\bibfield  {title} {\bibinfo {title} {Yang-lee edge singularity and $\phi^3$ field theory},\ }\href {https://doi.org/10.1103/PhysRevLett.40.1610} {\bibfield  {journal} {\bibinfo  {journal} {Physical Review Letters}\ }\textbf {\bibinfo {volume} {40}},\ \bibinfo {pages} {1610} (\bibinfo {year} {1978}{\natexlab{b}})}\BibitemShut {NoStop}%
\bibitem [{\citenamefont {Alves}\ \emph {et~al.}(2000)\citenamefont {Alves}, \citenamefont {Berg},\ and\ \citenamefont {Sanielevici}}]{alves2000}%
  \BibitemOpen
  \bibfield  {author} {\bibinfo {author} {\bibfnamefont {N.~A.}\ \bibnamefont {Alves}}, \bibinfo {author} {\bibfnamefont {B.~A.}\ \bibnamefont {Berg}},\ and\ \bibinfo {author} {\bibfnamefont {S.}~\bibnamefont {Sanielevici}},\ }\bibfield  {title} {\bibinfo {title} {Partition function zeros and phase transitions for finite lattice systems},\ }\href {https://doi.org/10.1016/S0550-3213(00)00381-0} {\bibfield  {journal} {\bibinfo  {journal} {Nuclear Physics B}\ }\textbf {\bibinfo {volume} {583}},\ \bibinfo {pages} {471} (\bibinfo {year} {2000})}\BibitemShut {NoStop}%
\bibitem [{\citenamefont {Glasser}\ and\ \citenamefont {Janke}(2001)}]{glasser2001}%
  \BibitemOpen
  \bibfield  {author} {\bibinfo {author} {\bibfnamefont {M.~L.}\ \bibnamefont {Glasser}}\ and\ \bibinfo {author} {\bibfnamefont {W.}~\bibnamefont {Janke}},\ }\bibfield  {title} {\bibinfo {title} {Complex temperature zeros of the potts model on the square lattice},\ }\href {https://doi.org/10.1088/0305-4470/34/30/303} {\bibfield  {journal} {\bibinfo  {journal} {Journal of Physics A: Mathematical and General}\ }\textbf {\bibinfo {volume} {34}},\ \bibinfo {pages} {6237} (\bibinfo {year} {2001})}\BibitemShut {NoStop}%
\bibitem [{\citenamefont {Itzykson}\ \emph {et~al.}(1983)\citenamefont {Itzykson}, \citenamefont {Pearson},\ and\ \citenamefont {Zuber}}]{itzykson1983}%
  \BibitemOpen
  \bibfield  {author} {\bibinfo {author} {\bibfnamefont {C.}~\bibnamefont {Itzykson}}, \bibinfo {author} {\bibfnamefont {R.~B.}\ \bibnamefont {Pearson}},\ and\ \bibinfo {author} {\bibfnamefont {J.~B.}\ \bibnamefont {Zuber}},\ }\bibfield  {title} {\bibinfo {title} {Distribution of zeros in ising and gauge models},\ }\href {https://doi.org/10.1016/0550-3213(83)90593-0} {\bibfield  {journal} {\bibinfo  {journal} {Nuclear Physics B}\ }\textbf {\bibinfo {volume} {220}},\ \bibinfo {pages} {415} (\bibinfo {year} {1983})}\BibitemShut {NoStop}%
\bibitem [{\citenamefont {Kenna}(2004)}]{kenna2004}%
  \BibitemOpen
  \bibfield  {author} {\bibinfo {author} {\bibfnamefont {R.}~\bibnamefont {Kenna}},\ }\bibfield  {title} {\bibinfo {title} {Universal scaling relations for logarithmic corrections},\ }\href {https://doi.org/10.1016/j.nuclphysb.2004.06.030} {\bibfield  {journal} {\bibinfo  {journal} {Nuclear Physics B}\ }\textbf {\bibinfo {volume} {691}},\ \bibinfo {pages} {292} (\bibinfo {year} {2004})}\BibitemShut {NoStop}%
\bibitem [{\citenamefont {Bena}\ \emph {et~al.}(2005)\citenamefont {Bena}, \citenamefont {Droz},\ and\ \citenamefont {Lipowski}}]{bena2005}%
  \BibitemOpen
  \bibfield  {author} {\bibinfo {author} {\bibfnamefont {I.}~\bibnamefont {Bena}}, \bibinfo {author} {\bibfnamefont {M.}~\bibnamefont {Droz}},\ and\ \bibinfo {author} {\bibfnamefont {A.}~\bibnamefont {Lipowski}},\ }\bibfield  {title} {\bibinfo {title} {Statistical mechanics of equilibrium and nonequilibrium phase transitions: The yang-lee formalism},\ }\href {https://doi.org/10.1142/S0217979205032759} {\bibfield  {journal} {\bibinfo  {journal} {International Journal of Modern Physics B}\ }\textbf {\bibinfo {volume} {19}},\ \bibinfo {pages} {4269} (\bibinfo {year} {2005})}\BibitemShut {NoStop}%
\bibitem [{\citenamefont {Kortman}\ and\ \citenamefont {Griffiths}(1971)}]{kortman1971}%
  \BibitemOpen
  \bibfield  {author} {\bibinfo {author} {\bibfnamefont {P.~J.}\ \bibnamefont {Kortman}}\ and\ \bibinfo {author} {\bibfnamefont {R.~B.}\ \bibnamefont {Griffiths}},\ }\bibfield  {title} {\bibinfo {title} {Density of zeros on the lee-yang circle for two ising ferromagnets},\ }\href {https://doi.org/10.1103/PhysRevLett.27.1439} {\bibfield  {journal} {\bibinfo  {journal} {Physical Review Letters}\ }\textbf {\bibinfo {volume} {27}},\ \bibinfo {pages} {1439} (\bibinfo {year} {1971})}\BibitemShut {NoStop}%
\bibitem [{\citenamefont {Gennes}\ and\ \citenamefont {Prost}(2023)}]{crystals}%
  \BibitemOpen
  \bibfield  {author} {\bibinfo {author} {\bibfnamefont {P.}~\bibnamefont {Gennes}}\ and\ \bibinfo {author} {\bibfnamefont {J.}~\bibnamefont {Prost}},\ }\href {https://doi.org/10.1093/oso/9780198520245.001.0001} {\emph {\bibinfo {title} {The Physics of Liquid Crystals}}}\ (\bibinfo {year} {2023})\BibitemShut {NoStop}%
\bibitem [{\citenamefont {Baxter}(1980)}]{baxter1980}%
  \BibitemOpen
  \bibfield  {author} {\bibinfo {author} {\bibfnamefont {R.~J.}\ \bibnamefont {Baxter}},\ }\bibfield  {title} {\bibinfo {title} {Hard hexagons: exact solution},\ }\href@noop {} {\bibfield  {journal} {\bibinfo  {journal} {Journal of Physics A: Mathematical and General}\ }\textbf {\bibinfo {volume} {13}},\ \bibinfo {pages} {L61} (\bibinfo {year} {1980})}\BibitemShut {NoStop}%
\bibitem [{\citenamefont {Baxter}(1982)}]{baxter1982}%
  \BibitemOpen
  \bibfield  {author} {\bibinfo {author} {\bibfnamefont {R.~J.}\ \bibnamefont {Baxter}},\ }\href@noop {} {\emph {\bibinfo {title} {Exactly Solved Models in Statistical Mechanics}}}\ (\bibinfo  {publisher} {Academic Press},\ \bibinfo {year} {1982})\BibitemShut {NoStop}%
\end{thebibliography}%

\end{document}